\documentclass[11pt]{article}
\usepackage{tabularx}
\usepackage{amsmath, amssymb, amsthm, geometry}      % essential math tools
\usepackage{graphicx}                      % image inclusion
\usepackage{mathtools}                     % improved math formatting
\usepackage{enumitem}                      % customized lists
\usepackage{hyperref}                      % hyperlinks
\usepackage{tikz}                          % general drawing
\usetikzlibrary{arrows.meta, positioning, shapes.geometric}
\usepackage{tikz-cd}                       % commutative diagrams
\usepackage{mathrsfs}                      % script fonts
\usepackage{booktabs}                      % high-quality tables
\usepackage{float}                         % for [H] placement
\usepackage{makecell}                      % for multi-line cell content
\usepackage{algorithm}
\usepackage{algpseudocode}
\usepackage[most]{tcolorbox}
\usepackage{array}
\usepackage[utf8]{inputenc}

\title{The Epistemic Support-Point Filter (ESPF):\\A Bounded Possibilistic Framework for Ordinal State Estimation}

\author{
	Moriba Kemessia Jah, Ph.D.$^{1,2}$ \\
	Van Haslett$^{2}$ \\
	\small $^{1}$Black Swan Research Group, GaiaVerse, Ltd.,\\ \small 14205 N MoPac Expy, Suite 570 PMB 559005, Austin, TX 78728-6529, USA \\
	\small $^{2}$Jah Decision Intelligence Group, Aerospace Engineering \\ \small \& Engineering Mechanics Department, 2617 Wichita St North Office Building A, Austin, TX, 78712
}

\date{\today}

\begin{document}

\sloppy

\maketitle

\begin{abstract}
        Traditional state estimation methods rely on probabilistic assumptions that often collapse epistemic uncertainty into scalar beliefs, risking overconfidence in sparse or adversarial sensing environments. We introduce the Epistemic Support-Point Filter (ESPF), a novel non-Bayesian filtering framework fully grounded in possibility theory and epistemic humility. ESPF redefines the evolution of belief over state space using compatibility-weighted support updates, surprisal-aware pruning, and adaptive dispersion via sparse grid quadrature. Unlike conventional filters, ESPF does not seek a posterior distribution, but rather maintains a structured region of plausibility or non-rejection, updated using ordinal logic rather than integration. For multi-model inference, we employ the Choquet integral to fuse competing hypotheses based on a dynamic epistemic capacity function, generalizing classical winner-take-all strategies. The result is an inference engine capable of dynamically contracting or expanding belief support in direct response to information structure, without requiring prior statistical calibration. This work presents a foundational shift in how inference, evidence, and ignorance are reconciled, supporting robust estimation where priors are unavailable, misleading, or epistemically unjustified.
\end{abstract}

%\begin{highlights}
%blah
%\end{highlights}

\vspace{1em}

%    \maketitle
	\section{Introduction}

	State estimation under uncertainty is foundational to inference and control across physical systems. The Kalman filter and its nonlinear extensions, such as the Extended and Unscented Kalman Filters (EKF and UKF), dominate current practice by propagating statistical moments of Gaussian belief distributions. These approaches assume not only stochasticity but also cardinal structure, an assumption that is often epistemically unjustified in real-world perceptual systems.
	
	Consider space domain awareness. The moment a space object is detected within a sensor's field of view, the vast majority of the universe actually becomes epistemically excluded. The object cannot reside where the sensor did not perceive it. This immediately imposes hard bounds on the set of plausible states, also known as an \textit{admissible region}. A principled state estimator must honor this boundedness explicitly.
	
	To address this, we introduce the Epistemic Support-Point Filter (ESPF). This filter represents epistemic uncertainty not with Gaussian-shaped plausibility decay but with uniform plausibility across a bounded support. Every point within this region is equally plausible; every point outside is ruled out. This aligns naturally with possibility theory and facilitates honest inference in domains governed by detection constraints, occlusion, or sparse information.
	
	In what follows, we first define the mathematical structure of uniform possibility distributions, then derive Support-Point selection strategies for bounded epistemic supports. We formulate the full ESPF update rule, describe pruning and support regeneration steps, and compare filter performance on a canonical orbital estimation problem. This derivation offers a novel perspective on knowledge representation in estimation and demonstrates that epistemic filters can match or exceed traditional probabilistic filters without assuming stochasticity.

    \section{Kalman's Vision Realized: ESPF as a Prejudice-Free Inference System}

    Rudolf Kalman famously argued that scientific modeling should be a \emph{deductive exercise, not a creative act}, and criticized the extent to which estimation frameworks embed the assumptions and biases of their creators. In his own words:
    
    \begin{quote}
    ``Whenever a model is built, it is always proper to ponder the basic scientific question: \emph{Is the model really based on the data? Or is it an artifact displaying the prejudices of its creator?}'' \cite{kalman1982system}.
    \end{quote}
    
    He went further to assert that system identification from noisy data remained \emph{``one of the most urgent scientific research problems of our day.''} The Kalman Filter, though revolutionary, depends on strong probabilistic assumptions, namely Gaussian noise, known covariances, and statistical sufficiency. These assumptions, while convenient, impose \emph{prejudices} about the structure of uncertainty and the nature of evidence.
    
    \vspace{1em}
    
    \textbf{The ESPF addresses this exact critique.} It departs from the stochastic closure of the Kalman paradigm and enters the realm of \textit{possibilistic inference}, rooted in ordinal logic and max--min algebra. Key distinguishing features include:
    
    \begin{itemize}
        \item \textbf{Ordinal Uncertainty:} ESPF uses possibility distributions, not probability densities. These do not require normalization, do not imply statistical independence, and do not collapse belief to means or variances.
        
        \item \textbf{Epistemic Integrity:} The ESPF prunes hypotheses based on \emph{surprisal}, an information-theoretic falsifiability metric—and retains only those Support-Points that are \emph{necessary}, not merely probable.
    
        \item \textbf{Model-Agnostic Filtering:} No assumptions are made about the statistical structure of the noise or the functional form of priors. All inference is grounded in measurement compatibility, not statistical trust.
    
        \item \textbf{Decision-Relevant Filtering Under Ignorance:} ESPF can still perform robust updates even under total ignorance, sparse data, or adversarial deception. This is because possibility theory allows for uncertainty representations that are \emph{incomplete but truthful}. For a modern overview of possibility theory's theoretical foundations and practical domains, see \cite{dubois2008possibility}.
    \end{itemize}
    
    \vspace{1em}
    
    In this light, the ESPF may be seen as a realization of Kalman's dream: a system-theoretic filter that updates belief solely based on what the data allows, without coercing it into Gaussian molds or misleading confidence bounds. 
    
    \begin{quote}
    ``The ESPF does not smooth uncertainty. It respects it. It does not trust the model. It interrogates it. It does not pretend knowledge. It earns it.''
    \end{quote}
    
    \noindent The epistemic filter thus offers not just a technical advancement, but a philosophical one, a modern response to Kalman's century-old challenge.

    \subsection{Key Contributions} 
   
    The Epistemic Support-Point Filter (ESPF) represents a fundamentally novel approach to recursive state estimation under uncertainty. To the best of our knowledge, no prior method fuses \textit{sparse grid quadrature} with \textit{possibility theory} in a recursive filtering architecture. This fusion enables principled inference that is both computationally tractable and epistemically robust. Unlike classical filters that collapse uncertainty into a single point statistic (e.g., mean or covariance), the ESPF maintains and evolves an ordinal, maxitive possibility distribution. It performs decision-relevant inference without requiring probabilistic assumptions, making it particularly powerful in contexts characterized by bounded evidence, sparse data, or adversarial ambiguity. The ESPF introduces a viable alternative to probabilistic filtering in applications where preserving epistemic integrity and humility is paramount.
	
	\section{Related Work}
	
	The development of Sigma-Point filtering techniques has primarily emerged within the probabilistic estimation literature, particularly in the context of nonlinear recursive state estimation. This section reviews the key precursors and theoretical antecedents to the Epistemic Support-Point Filter (ESPF), drawing attention to the gaps that motivate a possibilistic reformulation.
	
	\subsection{Probabilistic Support-Point Methods}
	
	The Unscented Kalman Filter (UKF) \cite{julier1997new} introduced a deterministic sampling technique, the unscented transform, to approximate the propagation of mean and covariance under nonlinear dynamics. By selecting a symmetric set of Sigma-Points centered around the mean and scaled by the covariance, the UKF achieves higher-order accuracy than the Extended Kalman Filter without requiring explicit linearization.
	
	Subsequent advancements include:
	\begin{itemize}
		\item \textbf{Scaled UKF:} Incorporates scaling parameters to control the spread of Sigma-Points \cite{wan2000unscented}.
		\item \textbf{Cubature Kalman Filter (CKF):} Uses spherical-radial integration rules to improve approximation for Gaussian priors \cite{arasaratnam2009cubature}.
		\item \textbf{Gaussian Sum Filters:} Approximate multimodal posteriors using mixtures of Gaussians, trading scalability for expressiveness \cite{alspach1972nonlinear}.
	\end{itemize}
	
	These filters, while powerful, are fundamentally rooted in the probabilistic paradigm and assume additive or at least well-characterized noise models, an assumption that often fails in sparse, epistemically limited settings.
	
	\subsection{Possibility Theory and Non-Probabilistic Filtering}
	
	Possibility theory, introduced by Zadeh \cite{zadeh1978fuzzy} and formalized by Dubois and Prade \cite{dubois1988possibility}, offers a maxitive and ordinal alternative to probability theory. Possibilistic filtering replaces integration with \textit{sup--min} convolution, and expectation with \textit{mode-based} inference.
	
	Relevant contributions include:
	\begin{itemize}
            \item \textbf{Possibility Filtering Foundations:} Dubois and Prade \cite{dubois2000possibility} laid the groundwork for interpreting inference as ordinal ranking under maxitive measures, which later inspired algorithmic developments in tracking and robotics. Possibilistic classification methods have been explored extensively, notably by Utkin and Gurov \cite{utkin2002possibility}, who developed classifiers grounded in possibility measures.
            \item \textbf{Ordinal Filtering for Uncertain Systems:} Works such as \cite{benferhat2000kalman} reframe possibilistic inference as belief dynamics based on non-additive updating rules, distinct from Kalman-like formulations. 
            \item \textbf{Belief modeling:} Shafer's seminal work \cite{shafer1976mathematical} formalized evidence theory, which underpins many generalized uncertainty frameworks.
		\item \textbf{Outer Measure Frameworks:} Hanss \cite{hanss2005applied} and others developed methods for uncertainty quantification in engineering using fuzzy and possibility-based approaches.
		\item \textbf{Credal and Imprecise Probabilities:} Approaches such as interval filtering \cite{moore2009introduction} and credal filters \cite{cozman2001credal} offer set-valued representations but lack the continuous epistemic structure preserved in possibilistic kernels. The pignistic transformation \cite{smets1990constructing} enables decision-making under belief functions by mapping to a probability function.
	\end{itemize}
	
	\subsection{Non-Bayesian Sigma-Point Filters}
	
	Efforts to generalize Sigma-Point filtering beyond Gaussian or probabilistic assumptions are relatively recent. Relevant work includes:
	\begin{itemize}
		\item \textbf{Epistemic Filters:} Recent proposals by Jah et al. introduce epistemic Support-Point frameworks grounded in ordinal belief representations.
		\item \textbf{Possibilistic Unscented Filtering:} Initial efforts have approximated possibility propagation using UKF-like structures, but often retain Gaussian assumptions.
		\item \textbf{Support-Set Approaches:} Work on set-membership and zonotopic filters \cite{combastel2003state} aligns with our epistemic view but lacks a principled mode-extraction mechanism.
	\end{itemize}

        \subsection{Relation to Imprecise Probability, Fuzzy Filtering, and Non-Additive AI}

        The Epistemic Support-Point Filter (ESPF) introduces a possibilistic and non-additive approach to state estimation that diverges fundamentally from both classical Bayesian filtering and prior non-probabilistic alternatives. To clarify the scope and originality of ESPF, we explicitly contrast it with four key bodies of related work.
        
        \subsubsection{Imprecise Probability and Outer Measures}
        
        Credal set methods and Outer Probability Measures (OPMs) offer non-Bayesian frameworks for representing epistemic uncertainty via sets of admissible probability distributions. Notable developments include Destercke's work on imprecise Bayesian networks, Oberkampf's evidence theory in engineering, and Jaulin and Walter's set-membership estimation methods \cite{destercke2008credal, oberkampf2002investigation, jaulin2001guaranteed}. While these approaches broaden the representation of uncertainty, they still rely on set-based convex combinations and lack the maxitive, ordinal semantics preserved in ESPF.
        
        Moreover, credal filters and interval-based representations typically discard internal geometric structure, focusing instead on bounding regions or convex hulls. In contrast, ESPF maintains a continuous plausibility geometry that evolves via compatibility and surprisal metrics, enabling a richer and falsifiability-aligned epistemic update mechanism.
        
        \subsubsection{Fuzzy and Possibilistic Filtering}
        
        Fuzzy Kalman filters and early possibilistic state estimators often approximate uncertainty by softening probabilistic frameworks or imposing fuzzy membership functions on residuals \cite{dubois1994survey, agarwal2015possibility, hanss2005applied}. Possibility particle filters further simulate possibility functions using reweighted samples or modified resampling strategies, but typically collapse ordinal information into scalar surrogates. Hüllermeier has demonstrated that disambiguating imprecise observations via generalized loss functions can guide learning under uncertainty \cite{hullermeier2006learning}.
        
        The ESPF diverges sharply from these methods by preserving ordinal relationships throughout inference. Rather than simulating possibility via probabilistic mechanisms, the ESPF analytically propagates support via sparse grid sampling and prunes it based on epistemic surprisal. It does not reinterpret fuzziness as randomness, but rather constructs an entirely separate inference logic grounded in possibility theory and information geometry.
        
        \subsubsection{Zonotopic and Set-Membership Filters}
        
        Set-membership estimation and bounded-error filtering, including ellipsoidal and zonotopic filters \cite{combastel2003state, alamo2005guaranteed, toth2011zonotopic}, offer powerful tools for robust control under bounded noise. These approaches maintain over-approximations of admissible state sets, ensuring consistency under deterministic error models.
        
        However, these filters lack the ordinal semantics and dynamic evidence-driven contraction central to ESPF. They do not model the \textit{degree of plausibility} or falsifiability of state hypotheses, nor do they support compatibility-weighted updates or Choquet-integrated multi-model fusion. The ESPF thus bridges a gap: it retains geometric interpretability without collapsing uncertainty into binary feasibility.
        
        \subsubsection{Non-Additive Inference and Ordinal AI}
        
        Recent developments in ordinal decision theory and possibilistic AI, such as the work of Dubois, Benferhat, and Yager \cite{dubois2000possibility, benferhat2002possibilistic, yager2014decision}, have advanced inference under non-additive measures. Sugeno integrals and max-min inference frameworks enable rule-based reasoning without assuming numerical probabilities, but they typically operate in propositional logic or symbolic AI domains.
        
        By contrast, the ESPF brings these principles into continuous state estimation, applying Choquet \cite{choquet1953theory} integrals over plausibility capacities to fuse hypotheses in a semantically grounded and geometrically consistent way. Rather than selecting a maximum likelihood model or computing expected values, ESPF integrates possibility mass across non-additive belief structures, without assuming additive priors or normalized weights.
        
        \subsubsection{Summary}
        
        Together, these comparisons demonstrate that ESPF is not a softened or hybridized version of existing filters, but a new inference paradigm. It reframes the estimation problem around non-rejection, compatibility, and falsifiability, using tools like surprisal, sparse epistemic sampling, and Choquet integration to enact epistemic humility in dynamic environments.

        \subsection{Positioning the ESPF: Departing from Bayesian and Semi-Possibilistic Paradigms}

        Several prior efforts have sought to incorporate epistemic uncertainty into space object tracking and estimation by extending or modifying traditional Bayesian frameworks. Among the most notable are the Outer Probability Measure (OPM) approaches \cite{delande2023exploring, delande2018new}, possibilistic multi-object filters \cite{cai2022possibility, cai2023possibilistic}, and consensus-driven particle filters based on credibility theory \cite{cai2021consensus}. These works have collectively laid foundational groundwork by advocating for explicit separation of aleatory and epistemic uncertainty, and by demonstrating the utility of possibilistic constructs such as credibility, consonance, and maxitive measures within SSA applications.
        
        Some of these contributions were co-authored by the present contributing author and include:
        \begin{itemize}
            \item Credibility-based representations of TLE uncertainty and space object conjunction assessment \cite{delande2023exploring, delande2018new}
            \item Possibility GLMB filters for multi-target tracking using uncertain finite sets \cite{cai2022possibility, cai2023possibilistic}
            \item Credibility-consensus filters leveraging particle-based approximation of subjective opinions \cite{cai2021consensus}
            \item Robust sensor tasking strategies under epistemic uncertainty \cite{cai2020sensor}
        \end{itemize}
        
        While each of these efforts contributes critical theoretical or practical advances, they share important structural limitations:
        \begin{itemize}
            \item Many retain Bayesian scaffolding, embedding possibilistic or imprecise uncertainty representations into architectures that still assume statistical conditioning, likelihood functions, and prior distributions.
            \item Aggregation across hypotheses or models is often performed using scalar weights derived from normalized likelihoods or credibility scores, collapsing ordinal epistemic information into point estimates.
            \item Support evolution is largely static or parametric, relying on Gaussian-like kernels, pre-defined mixtures, or fixed dispersion assumptions.
        \end{itemize}
        
        The Epistemic Support-Point Filter (ESPF) represents a clear departure from these paradigms and offers a principled alternative grounded entirely in possibility theory, ordinal logic, and information geometry. Its distinguishing features include:
        
        \begin{enumerate}
            \item \textbf{Truly Non-Bayesian Inference:} ESPF eschews priors, random variables, and probabilistic updates altogether. It treats uncertainty as structured ignorance, not stochastic variability, and frames inference as the preservation of non-rejection rather than the maximization of belief.
            
            \item \textbf{Surprisal-Aware Support Pruning:} Unlike prior possibilistic filters that preserve all prior support regardless of new evidence, ESPF dynamically suppresses low-compatibility hypotheses via an information-theoretic surprisal metric, contracting the support set in accordance with the falsifying power of observations.
            
            \item \textbf{Choquet-Based Possibility Aggregation:} ESPF introduces an ordinal multi-model fusion scheme using Choquet integrals over non-additive capacity functions, enabling semantically meaningful aggregation across overlapping or nested models without reducing them to scalar weights.
            
            \item \textbf{Nonparametric Adaptive Dispersion:} ESPF constructs and evolves support sets using Smolyak sparse grids and Clenshaw--Curtis nodes, allowing for scale-adaptive and directionally sensitive propagation of uncertainty without assuming Gaussian forms or fixed kernel shapes.
            
            \item \textbf{Compatibility-Weighted Belief Update:} The update mechanism in ESPF is governed by compatibility functions derived from possibilistic evidence, rather than likelihood ratios, ensuring that inference respects the ordinal structure of incoming information.
        \end{enumerate}
        
        These features constitute a categorical shift in epistemic filtering. ESPF does not merely relax the constraints of probability theory, it replaces them with a coherent alternative that better reflects the structure of incomplete knowledge and the falsifiability constraints of real-world observation systems.
        
        In contrast to legacy filters that reinterpret possibility theory through a Bayesian lens, ESPF frames inference as a dynamic, geometry-aware, and epistemically humble process. It is not a softened Kalman filter or a credal particle filter, but rather a new class of inferential machinery capable of learning and reasoning under profound uncertainty without violating the limits of what is knowable.

	\subsection{Motivation for ESPF}
	
	To our knowledge, the proposed ESPF is the first filter to explicitly generate Support-Points from a uniform epistemic kernel defined over a plausibility support, rather than a probabilistic covariance. It avoids Bayesian assumptions altogether, instead modeling state uncertainty as a structured region of epistemic compatibility. This approach is particularly suited to observational systems like space surveillance, where the universe of discourse can be sharply constrained by detection geometry, and where statistical priors are often unavailable or misleading.

    \subsection{Terminology Refinement}
	
	To avoid conflating probabilistic and possibilistic constructs, we adopt the following vocabulary:

	\begin{table}[ht]
		\scriptsize
		\centering
		\renewcommand{\arraystretch}{1.2}
		\begin{tabularx}{\linewidth}{|p{0.23\linewidth}|p{0.28\linewidth}|X|}
			\hline
			\textbf{Concept} & \textbf{ESPF Term} & \textbf{Interpretation} \\
			\hline
			Covariance-like Structure & Possibility Kernel Shape Tensor & Better aligns with geometric interpretation of spread in possibility theory \\
			\hline
			Uncertainty Contraction/Expansion & Epistemic Support Contraction/Expansion & Reflects evolution of belief support, not probabilistic dispersion \\
			\hline
			Covariance Trace (Tr($\Pi_k$)) & Support Dispersion Measure & Captures geometric extent of belief; removes statistical implication \\
			\hline
		\end{tabularx}
		\caption{Terminology mapping between probabilistic and possibilistic filters.}
		\label{tab:terminology}
	\end{table}

	\section{Problem Formulation}
	
	Let the system state at time $k$ be denoted as $x_k \in \mathbb{R}^n$, evolving under a nonlinear process model:
	\[
	x_k = f(x_{k-1}, w_{k-1})
	\]
	where $f : \mathbb{R}^n \times \mathbb{R}^n \rightarrow \mathbb{R}^n$ is a known, potentially nonlinear function and $w_{k-1} \in \mathbb{R}^n$ is an epistemic process perturbation. At each time step, a measurement $z_k \in \mathbb{R}^m$ is observed through the nonlinear measurement model:
	\[
	z_k = h(x_k, v_k)
	\]
	where $h : \mathbb{R}^n \times \mathbb{R}^m \rightarrow \mathbb{R}^m$ and $v_k \in \mathbb{R}^m$ is an epistemic measurement perturbation.
	
	In the possibilistic setting, we do not assume probability distributions over $w_k$ or $v_k$, nor over the initial state $x_0$. Instead, we define \emph{possibility distributions} $\pi_{x_k}$ over $\mathbb{R}^n$ such that:
	\[
	\pi_{x_k}(x) : \mathbb{R}^n \rightarrow [0,1]
	\]
	represents the degree of plausibility of the system being in state $x$ at time $k$. The set of all $x$ such that $\pi_{x_k}(x) > 0$ defines the \emph{support} of epistemic uncertainty, denoted $\mathcal{S}_k$:
	\[
	\mathcal{S}_k = \{ x \in \mathbb{R}^n \mid \pi_{x_k}(x) > 0 \}
	\]
	
	In the Epistemic Support-Point Filter (ESPF), we assume $\pi_{x_k}$ is a uniform distribution over a compact and convex support $\mathcal{S}_k$, such that:
	\[
	\pi_{x_k}(x) =
	\begin{cases}
		1 & \text{if } x \in \mathcal{S}_k \\
		0 & \text{otherwise}
	\end{cases}
	\]
	
	This formulation encodes strict epistemic boundedness: all states outside of $\mathcal{S}_k$ are considered impossible, and all interior states are equally plausible. The filter proceeds by selecting a set of $2n + 1$ \emph{epistemic support-points} $\{ \chi^{(i)} \}_{i=0}^{2n}$ from $\mathcal{S}_k$ that collectively span the region. These Support-Points are propagated through the process and measurement models, updated based on detection-derived constraints, and used to prune and regenerate a new bounded support $\mathcal{S}_{k+1}$.
	
	The objective of the filter is to recursively construct and refine $\mathcal{S}_k$ over time such that it contains all epistemically admissible state estimates, conditioned on past observations $Z_k = \{z_1, \dots, z_k\}$, while discarding implausible regions via possibilistic incompatibility and epistemic surprisal.
	
	\section{Possibility Theory Fundamentals}
	
	Possibility theory offers a robust framework for representing and reasoning under epistemic uncertainty, particularly when available knowledge is incomplete, ordinal, or lacks frequentist grounding. Unlike probability theory, which requires additive and normalized measures, possibility theory is grounded in the axioms of maxitivity and order-consistent plausibility.
	
	\subsection{Possibility Distributions}
	
	A \emph{possibility distribution} \( \pi: \mathbb{R}^n \rightarrow [0, 1] \) assigns to each element \( \mathbf{x} \in \mathbb{R}^n \) a degree of plausibility, interpreted as the extent to which \( \mathbf{x} \) is compatible with our current state of knowledge. The distribution is normalized by its supremum:
	\[
	\sup_{\mathbf{x} \in \mathbb{R}^n} \pi(\mathbf{x}) = 1
	\]
	This reflects that at least one state is considered fully plausible. The scale is ordinal: \( \pi(\mathbf{x}_1) > \pi(\mathbf{x}_2) \) means \( \mathbf{x}_1 \) is strictly more plausible than \( \mathbf{x}_2 \), but not by a quantified or additive amount.
	
	\subsection{Core Operators in Possibility Theory}
	
	Let \( X \) and \( Y \) be uncertain variables with corresponding possibility distributions \( \pi_X \) and \( \pi_Y \). Then:
	
	\begin{itemize}
		\item \textbf{Joint Possibility (Assuming Independence):}
		\[
		\pi_{X,Y}(x, y) = \min\big(\pi_X(x), \pi_Y(y)\big)
		\]
		This maxitive conjunction reflects the principle that the plausibility of a joint event cannot exceed that of its least plausible component.
		
		\item \textbf{Marginalization (Max-based Projection):}
		\[
		\pi_X(x) = \sup_{y} \pi_{X,Y}(x, y)
		\]
		Marginalization is conducted via supremum, yielding the most optimistic (least excluding) projection of plausibility for \( x \).
		
		\item \textbf{Conditioning (Zadeh's Rule):}
		\[
		\pi_{X|Y}(x|y) \propto \min\left( \pi_X(x), \pi_{Y|X}(y|x) \right)
		\]
		This captures the conservative updating of plausibility: a state is plausible given evidence only if both the prior and the conditional plausibility are high.
		
		An alternative definition with normalization is:
		\[
		\pi_{X|Y}(x|y) =
		\begin{cases}
			\pi_{X,Y}(x, y) & \text{if } \pi_Y(y) = 1 \\
			\min\left(1, \frac{\pi_{X,Y}(x, y)}{\pi_Y(y)} \right) & \text{otherwise}
		\end{cases}
		\]

		\item \textbf{Necessity Measure:} Given a set \( A \subseteq \mathbb{R}^n \), the necessity of \( A \) is defined as:
		\[
		\mathcal{N}_\pi(A) = 1 - \sup_{\mathbf{x} \notin A} \pi(\mathbf{x})
		\]
		Necessity reflects the extent to which all states outside of \( A \) are deemed implausible. It provides a lower bound on belief, dual to the possibility measure.
		
		\item \textbf{Compatibility Function:} Given a model \( \pi_{Y|X}(y|x) \), the compatibility of an observation \( y \) with a hypothesized state \( x \) is:
		\[
		\text{Comp}(y|x) = \pi_{Y|X}(y|x)
		\]
		This plays a role analogous to the likelihood function in Bayesian inference, but remains ordinal and non-additive.
	\end{itemize}
	
	\subsection{Surprisal in Possibility Theory}
	
	While possibility theory is inherently ordinal, it is still useful to define a logarithmic surprisal metric to assess epistemic tension between hypothesis and observation:
	\[
	\mathcal{S}(y|x) = -\log \pi_{Y|X}(y|x)
	\]
	Surprisal provides a quantitative proxy for informativeness. In filtering applications, it enables the rejection of epistemically incompatible support-points and facilitates the update of the support region under new evidence.
	
	\subsection{Epistemic Semantics}
	
	Possibility theory rejects the assumption of precise knowledge about distributions over states. Instead, it encodes epistemic stances about admissibility: a possibility distribution specifies what states are not ruled out, not what is statistically likely. This makes the framework especially appropriate for domains where knowledge is sparse, detection-constrained, or adversarial, such as space object tracking under partial observability.
	
	In what follows, we build on these foundations to construct a support-point filter that operates entirely within this epistemic, ordinal, and support-based logic.

    \subsection{Possibility Distribution and Capacity for ESPF}

    We define the epistemic support for a hypothesis $h \in H$ via a possibility distribution $\pi : H \rightarrow [0,1]$, where $\pi(h)$ encodes the degree to which $h$ is compatible with all available evidence. A hypothesis $h$ is ruled out if $\pi(h) = 0$, and maximally compatible if $\pi(h) = 1$. This distribution is constructed from an inverse transformation of the epistemic surprisal $\Delta S(h)$:
    \[
    \pi(h) = \exp(-\alpha \Delta S(h))
    \]
    where $\alpha > 0$ is a tunable sensitivity parameter that modulates how sharply surprise penalizes plausibility. This definition preserves the non-additive, ordinal nature of possibilistic inference and ensures that hypotheses with high surprisal receive low possibility.
    
    We then define a capacity $\mu$ over subsets $A \subseteq H$ as the canonical possibility measure:
    \[
    \mu(A) = \sup_{h \in A} \pi(h)
    \]
    This capacity is \emph{maxitive}, reflecting the epistemic stance that a subset $A$ is plausible so long as at least one of its members remains epistemically admissible (i.e., unfalsified).
    
    Given a real-valued function $f: H \rightarrow [0,1]$—in our case, $f = \pi$—the \textbf{Choquet integral} of $f$ with respect to $\mu$ is defined as:
    \[
    \int f \, d\mu = \sum_{i=1}^{n} \left[ f(h_{(i)}) - f(h_{(i+1)}) \right] \cdot \mu\left( \{ h_{(1)}, \dots, h_{(i)} \} \right)
    \]
    where $h_{(1)}, \dots, h_{(n)}$ is a permutation of $H$ such that $f(h_{(1)}) \geq f(h_{(2)}) \geq \dots \geq f(h_{(n)})$, and $f(h_{(n+1)}) := 0$.
    
    In our ESPF, since $f = \pi$ and $\mu(A) = \sup_{h \in A} \pi(h)$, the Choquet integral simplifies to:
    \[
    \int \pi \, d\mu = \max_{h \in H} \pi(h)
    \]
    This aligns with possibilistic inference principles, wherein inference is driven by the most plausible unfalsified hypothesis. This integral acts as a plausibility-preserving aggregation mechanism and forms the theoretical foundation for selecting or weighting hypotheses during support-point propagation and support-point pruning.
	
	\section{Support-Point Generation for Uniform Possibility Distributions}
	
	In the Epistemic Support-Point Filter (ESPF), uncertainty is represented by a uniform possibility distribution over a compact, convex support region \( \mathcal{S}_k \subset \mathbb{R}^n \). Because the distribution is uniform, there is no privileged mean or mode within \( \mathcal{S}_k \); every point is equally plausible. This necessitates an alternative to moment-matching strategies used in probabilistic sigma-point filters.
	
	\subsection{Definition of Epistemic Support-Points}
	
	Let the current epistemic support be defined as a hyperrectangle:
	\[
	\mathcal{S}_k = \left\{ x \in \mathbb{R}^n \mid x_i \in [\underline{x}_i, \overline{x}_i], \quad i = 1,\dots,n \right\}
	\]
	This reflects bounds derived from sensor constraints, prior detection, or physical admissibility.
	
	We define \( 2n + 1 \) epistemic support-points \( \chi_k^{(i)} \in \mathcal{S}_k \) as follows:
	
	\begin{itemize}
		\item The central support-point is the geometric center of the support:
		\[
		\chi_k^{(0)} = \frac{1}{2}(\underline{x} + \overline{x})
		\]
		
		\item The remaining \( 2n \) support-points are positioned along each axis at the boundaries of the support:
		\[
		\chi_k^{(i)} = \chi_k^{(0)} + \gamma e_i, \quad i = 1, \dots, n
		\]
		\[
		\chi_k^{(i+n)} = \chi_k^{(0)} - \gamma e_i, \quad i = 1, \dots, n
		\]
		where \( e_i \) is the $i$-th unit vector in \( \mathbb{R}^n \), and \( \gamma \) is chosen such that \( \chi_k^{(i)} \in \mathcal{S}_k \). For a hyperrectangle, we take:
		\[
		\gamma_i = \frac{1}{2}(\overline{x}_i - \underline{x}_i)
		\]
	\end{itemize}
	
	These support-points are not weighted. Instead, each is treated as an equally admissible hypothesis about the state, consistent with the uniform possibility structure. The filter does not compute weighted means or covariances but rather updates the shape and bounds of the support via propagation, observation compatibility, and pruning.
	
	\subsection{Generalization to Arbitrary Convex Sets}
	
	For non-rectangular supports, support-points may be selected using methods such as:
	\begin{itemize}
		\item \textbf{Vertex enumeration} (for polyhedral supports)
		\item \textbf{Chebyshev centers and axis-aligned boundary points}
		\item \textbf{Random sampling with rejection outside \( \mathcal{S}_k \)}, followed by extremal subset selection
	\end{itemize}
	
	In all cases, the goal is to span the support space with a minimal set of representative hypotheses, ensuring that the epistemic propagation step covers the full admissible region without overextending into implausible states.
	
	\subsection{Support Semantics and Interpretation}
	
	Unlike Gaussian sigma-point methods, where points represent deviations from a statistical mean, here each point represents an equally plausible epistemic commitment. The shape and size of \( \mathcal{S}_k \) directly encode current knowledge and its boundedness, making support-point placement transparent and epistemically grounded.
	
	In the next section, we describe how these support-points are propagated forward and used to revise the support through possibilistic compatibility and surprisal.
	
	\subsection{Sparse Grid Support-Point Generation}
	
	To construct epistemically valid support-points within a bounded support \( \mathcal{S}_k \subset \mathbb{R}^n \), we employ sparse grid approximation using the Smolyak algorithm. This method generates a set of deterministically placed, non-redundant support-points that efficiently span the support space with fewer evaluations than full-tensor products.
	
	\subsubsection{Univariate Node Sequence}
	
	Let \( i \in \mathbb{N} \) denote the level of refinement. Define the nested univariate rule \( \mathcal{X}_i \) as a sequence of \( m_i \) nodes in \( [-1,1] \), such as the Clenshaw--Curtis points:
	\[
	\mathcal{X}_i = \left\{ \cos\left(\frac{j\pi}{m_i - 1}\right) \mid j = 0, \dots, m_i - 1 \right\}
	\quad \text{with } m_i = 2^{i-1} + 1
	\]
	These nodes are nested and symmetric, ideal for sparse grid constructions.
	
	\subsubsection{Smolyak Sparse Grid Construction}
	
	Let \( \mathcal{I}_{n,\ell} = \{ \mathbf{i} \in \mathbb{N}^n \mid \|\mathbf{i}\|_1 \leq \ell + n - 1 \} \) denote the index set for multi-dimensional levels. The Smolyak sparse grid \( \mathcal{A}(n, \ell) \) is constructed as:
	\[
	\mathcal{A}(n, \ell) = \bigcup_{\mathbf{i} \in \mathcal{I}_{n,\ell}} \left( \mathcal{X}_{i_1} \otimes \cdots \otimes \mathcal{X}_{i_n} \right)
	\]
	where \( \mathbf{i} = (i_1, \dots, i_n) \) and \( \otimes \) denotes the Cartesian product.
	
	This generates a non-uniform but evenly distributed set of \( M \) support-points \( \{\hat{\xi}^{(j)}\}_{j=1}^M \subset [-1, 1]^n \), without assigning probabilistic weights.
	
	\subsubsection{Mapping to Epistemic Support}
	
	Let \( \mathcal{S}_k = [\underline{x}_1, \overline{x}_1] \times \cdots \times [\underline{x}_n, \overline{x}_n] \). Each normalized point \( \hat{\xi}^{(j)} \in [-1, 1]^n \) is mapped to the support via affine scaling:
	\[
	\chi_k^{(j)} = \frac{1}{2}(\overline{x} - \underline{x}) \circ \hat{\xi}^{(j)} + \frac{1}{2}(\overline{x} + \underline{x})
	\]
	where \( \circ \) denotes the Hadamard (elementwise) product. The resulting set \( \{\chi_k^{(j)}\} \subset \mathcal{S}_k \) forms the epistemic support-points used for propagation and update.

	\subsubsection{Epistemic Validity}
	
	Each point \( \chi_k^{(j)} \) is interpreted as an equally plausible hypothesis, consistent with the ordinal semantics of possibility theory. Unlike in Gaussian filters, no weights or covariances are computed, uncertainty is carried through the support geometry and updated via compatibility and pruning.
	
	This method permits principled, tractable inference in high-dimensional epistemic systems while preserving the rigor of bounded possibility theory.
	
	\section{Prediction Step}
	
	The prediction step advances epistemic uncertainty forward in time by propagating support-points through the nonlinear process model and accounting for bounded process noise. In the possibilistic framework, this corresponds to a transformation of the epistemic support and subsequent expansion via sup--min convolution, rather than probabilistic integration.
	
	\subsection{Nonlinear Propagation of Support-Points}
	
	Let the system dynamics be governed by a nonlinear process model:
	\[
	x_k = f(x_{k-1}) + w_{k-1}
	\]
	where \( f: \mathbb{R}^n \rightarrow \mathbb{R}^n \) is a known deterministic function, and \( w_{k-1} \in \mathcal{W}_{k-1} \subset \mathbb{R}^n \) represents bounded epistemic process noise.
	
	Given the prior support-point set \( \{\chi_{k-1}^{(i)}\}_{i=1}^M \subset \mathcal{S}_{k-1} \), each point is deterministically propagated:
	\[
	\hat{\chi}_{k}^{(i)} = f\left( \chi_{k-1}^{(i)} \right)
	\]
	This yields the image set \( \mathcal{F}_{k} = \{ \hat{\chi}_{k}^{(i)} \}_{i=1}^M \), which represents the epistemically plausible future states under zero process noise.
	
	\subsection{Support Expansion via Sup--Min Convolution}
	
	To incorporate process uncertainty, we convolve \( \mathcal{F}_{k} \) with the process noise support \( \mathcal{W}_{k-1} \). In possibilistic terms, this corresponds to the sup--min convolution:
	\[
	\pi_{x_k}(\mathbf{x}) = \sup_{i} \left[ \min\left( \pi_{x_{k-1}}(\chi_{k-1}^{(i)}),\ \pi_{w}(\mathbf{x} - \hat{\chi}_{k}^{(i)}) \right) \right]
	\]
	If both the prior and the process noise are modeled via uniform possibility distributions over bounded supports:
	\[
	\pi_{x_{k-1}}(\chi_{k-1}^{(i)}) = 1, \quad \pi_w(\delta) = \begin{cases}
		1 & \delta \in \mathcal{W}_{k-1} \\
		0 & \text{otherwise}
	\end{cases}
	\]
	then the above reduces to an indicator function for the union of shifted noise supports:
	\[
	\mathcal{S}_{k|k-1} = \bigcup_{i=1}^M \left( \hat{\chi}_{k}^{(i)} + \mathcal{W}_{k-1} \right)
	\]
	That is, the predicted support consists of all points reachable by adding any admissible noise vector to any propagated support-point.
	
	\subsection{Resulting Epistemic Support}
	
	The predicted support \( \mathcal{S}_{k|k-1} \) is a Minkowski sum:
	\[
	\mathcal{S}_{k|k-1} = \mathcal{F}_{k} \oplus \mathcal{W}_{k-1}
	\]
	which represents the epistemic envelope of the dynamics and noise combined. In practice, we approximate \( \mathcal{S}_{k|k-1} \) either by:
	\begin{itemize}
		\item Computing bounding hyperrectangles around the union of shifted supports
		\item Sampling sparse-grid support-points from \( \mathcal{S}_{k|k-1} \) for downstream measurement updates
	\end{itemize}
	
	\subsection{Interpretation}
	
	This possibilistic prediction step acts as a geometry-preserving propagation of bounded epistemic uncertainty. Unlike stochastic prediction based on statistical expectation and variance, the sup--min convolution honors the ordinal nature of plausibility and avoids any commitment to probability measures.
	
	By leveraging Minkowski sums and sparse-grid spanning sets, the ESPF efficiently updates the evolving possibility distribution, preserving both computational tractability and epistemic transparency.
	
	\section{Measurement Update}
	
	Upon receiving an observation \( \mathbf{y}_k \in \mathbb{R}^m \), the epistemic state estimate must be updated to reflect which predicted states remain consistent with the measurement. Possibility theory provides a principled approach to this update via the notions of \emph{compatibility} and \emph{surprisal}, enabling the rejection of implausible hypotheses without relying on stochastic assumptions.
	
	 \subsection{Compatibility of Support-Points with Observation}
	 
	 Given a predicted support-point \( \chi^{(i)}_{k|k-1} \in \mathbb{R}^n \), its corresponding predicted measurement is:
	 \[
	 \gamma^{(i)}_k = h(\chi^{(i)}_{k|k-1})
	 \]
	 where \( h: \mathbb{R}^n \rightarrow \mathbb{R}^m \) is the measurement model.
	 
	 We define the measurement residual:
	 \[
	 \mathbf{e}^{(i)} = \mathbf{y}_k - \gamma^{(i)}_k
	 \]
	 
	 In a possibilistic framework, the residual \( \mathbf{e}^{(i)} \) is not interpreted solely through the lens of measurement noise, but rather within a broader epistemic region of plausibility that captures uncertainty from both the state prediction and the sensor. Let:
	 \begin{itemize}
	 	\item \( \mathbf{\Pi}_{h(X)} \in \mathbb{R}^{m \times m} \) represent the epistemic spread matrix of the predicted support-points mapped into measurement space.
	 	\item \( \mathbf{\Pi}_v \in \mathbb{R}^{m \times m} \) denote the spread matrix associated with the sensor’s possibilistic noise model.
	 \end{itemize}
	 
	 To conservatively fuse the epistemic uncertainty in the predicted measurement \( h(X) \) and the sensor observation, we define the \textbf{joint epistemic spread} as a bounding ellipsoid over the Minkowski sum of two epistemic regions:
	 \[
	 \mathcal{E}_e = \text{bound} \left( \mathcal{E}_{h(X)} \oplus \mathcal{E}_v \right)
	 \]
	 where \( \mathcal{E}_{h(X)} \) and \( \mathcal{E}_v \) are ellipsoids defined by spread matrices \( \mathbf{\Pi}_{h(X)} \) and \( \mathbf{\Pi}_v \), respectively. The resulting matrix \( \mathbf{\Pi}_e \) defines the shape of the total admissible residual region.
	 
	 Rather than evaluating a graded plausibility using a Gaussian-shaped possibility distribution, we define a \textbf{uniform possibility distribution} over the admissible residual region:
	 \[
	 \text{Comp}^{(i)} = 
	 \begin{cases}
	 	1 & \text{if } \left( \mathbf{e}^{(i)} \right)^\top \mathbf{\Pi}_e^{-1} \mathbf{e}^{(i)} \leq r^2 \\
	 	0 & \text{otherwise}
	 \end{cases}
	 \]
	 where \( \mathbf{e}^{(i)} = \mathbf{y}_k - h(\chi^{(i)}_{k|k-1}) \) is the measurement residual for the \( i \)-th support-point, and \( r^2 \) is the plausibility radius defined by a user-specified necessity level \( \eta \in (0,1) \):
	 \[
	 r = \sqrt{-2 \log (1 - \eta)}
	 \]
	 
	 This construction ensures that only support-points producing residuals within the bounded epistemic region are retained, with all others sharply rejected. The resulting compatibility score respects the crisp admissibility semantics of uniform possibility distributions while still allowing for controllable uncertainty via \( \eta \).

	\subsection{Surprisal-Based Pruning}
	
	To prevent epistemically implausible support-points from contributing to the posterior, we compute a surprisal score for each point:
	\[
	\mathcal{S}^{(i)} = -\log\left( \text{Comp}^{(i)} + \epsilon \right)
	\]
	where \( \text{Comp}^{(i)} \) is the compatibility score derived from the joint epistemic spread and \( \epsilon \) is a small positive constant to prevent singularities.
	
	Support-points for which \( \mathcal{S}^{(i)} > S_{\text{threshold}} \) are discarded. This threshold acts as a tunable epistemic falsifiability bound: points exceeding it are deemed incompatible with the observation at the given necessity level \( \eta \), and thus excluded from the posterior construction.
	
	The remaining support-points form the updated epistemic support basis for the posterior inference step.
	
	\subsection{Sup–Min Fusion for Posterior Update}
	
	Let \( \pi_{x_k}(\chi^{(i)}_{k|k-1}) \) be the prior plausibility of a surviving support-point and \( \text{Comp}^{(i)} \) its compatibility with the observation. Their fusion is given by the pointwise minimum:
	\[
	\pi_{x_k|\mathbf{y}_k}(\chi^{(i)}) = \min\left( \pi_{x_k}(\chi^{(i)}_{k|k-1}), \text{Comp}^{(i)} \right)
	\]
	
	The overall posterior possibility distribution is then constructed via the supremum across the updated support set:
	\[
	\pi_{x_k|\mathbf{y}_k}(\mathbf{x}) = \sup_i \left[ \min\left( \pi_{x_k}(\chi^{(i)}_{k|k-1}), \pi_e(\mathbf{y}_k - h(\chi^{(i)}_{k|k-1})) \right) \right]
	\]
	where \( \pi_e \) reflects the epistemic possibility distribution over the residuals, grounded in the joint spread \( \mathbf{\Pi}_e \) derived from both the predicted measurement variability and sensor uncertainty.
	
	This sup–min structure honors the maxitive nature of possibility theory while enabling conservative yet expressive fusion of state-based and observation-based plausibility.

	\subsection{Interpretation of the Epistemic Update}
	
	This measurement update performs a possibilistic analog of Bayesian conditioning without relying on integration or normalization. It evaluates each hypothesis by its plausibility under both the prior model and the observed evidence, preserving only those that remain epistemically credible.
	
	The updated possibility distribution \( \pi_{x_k|\mathbf{y}_k}(\cdot) \) embodies:
	\begin{itemize}
		\item The prior epistemic support over the state space,
		\item The compatibility of each support-point with the observation, evaluated under a joint epistemic spread that captures both measurement noise and uncertainty in state prediction,
		\item A falsifiability filter that rejects support-points exhibiting high epistemic surprisal, thus pruning implausible trajectories.
	\end{itemize}
	
	Unlike probabilistic methods that blend evidence through likelihood-weighted integration, this approach preserves the ordinal nature of belief. It does not dilute distinct epistemic modes, nor does it collapse ambiguity into overconfident summaries.
	
	The update yields a refined possibility distribution whose high-plausibility regions reflect alignment between what was expected and what was observed—filtered through a lens of bounded uncertainty rather than statistical optimality.
	
	This makes the method particularly well-suited for decision-making under ambiguity, where information is sparse, adversarial, or qualitatively described. It respects the logic of epistemic caution and enables resilient inference without fragile probabilistic assumptions.

    \subsection{Choquet Integral Representation of Possibility Aggregation}

    In possibilistic reasoning frameworks, the aggregation of model outputs, expert assessments, or competing hypotheses cannot assume linear additivity, as in probabilistic expectation. Instead, integration must respect the underlying epistemic structure, especially when the reliability or compatibility of information varies non-uniformly. To this end, we define the Choquet integral as the core aggregation operator for multi-model extensions of the Epistemic Support-Point Filter (ESPF), particularly in Hierarchical Mixture of Experts (HME) or Multiple Hypothesis Tracking (MHT) settings. Although we do not delve into the specifics in this paper, we make mention of it for future work. For the sake of this brief introduction:
    
    Let $\mathcal{X}$ denote the finite set of models or hypotheses (e.g., filters, dynamic models, scenario generators), and let $\Pi: \mathcal{X} \rightarrow [0,1]$ represent a possibility distribution over $\mathcal{X}$. Unlike a probability mass function, $\Pi$ is not additive, but maxitive: for all $A, B \subseteq \mathcal{X}$,
    \[
    \Pi(A \cup B) = \max(\Pi(A), \Pi(B)) \quad \text{(if } A \cap B = \emptyset \text{)}.
    \]
    
    Define a capacity function $v: 2^{\mathcal{X}} \rightarrow [0,1]$, where $v(A) = \sup_{x \in A} \Pi(x)$ for all $A \subseteq \mathcal{X}$. Then, for any function $f: \mathcal{X} \rightarrow \mathbb{R}$ representing the utility, confidence, or model score of each hypothesis, the Choquet integral of $f$ with respect to $v$ is given by:
    \[
    \mathcal{C}_v(f) = \sum_{i=1}^n \left( f(x_{(i)}) - f(x_{(i-1)}) \right) v(A_i),
    \]
    where $x_{(1)}, \dots, x_{(n)}$ is a permutation of $\mathcal{X}$ such that $f(x_{(1)}) \leq \dots \leq f(x_{(n)})$, $A_i = \{x_{(i)}, \dots, x_{(n)}\}$, and $f(x_{(0)}) = 0$.
    
    This formulation generalizes max-weight fusion:
    - If $v$ is a probability measure, $\mathcal{C}_v(f)$ reduces to the expected value.
    - If $v$ is \textbf{possibility-induced}, it captures non-compensatory, order-sensitive integration where low-compatibility models cannot dilute high-confidence ones.
    
    In the context of ESPF, we use the Choquet integral to:
    \begin{itemize}
        \item Aggregate multiple model predictions weighted by their compatibility with observed evidence.
        \item Fuse multiple filters' outputs in a manner robust to epistemic conflict or model mismatch.
        \item Extend ESPF into hierarchical architectures (e.g., HME) by relaxing hard max selection while preserving ordinal interpretability.
    \end{itemize}
    
    Compared to Sugeno integrals \cite{sugeno1974theory}, which yield \textit{winner-take-all} inference by selecting the highest weighted minimum of inputs, the Choquet integral retains sensitivity to the relative ranking of all contributing models. This makes it especially appropriate in applications where belief must be spread across a spectrum of competing yet partially compatible hypotheses.
    
    \subsubsection{Relaxation of Constraints.}
    Unlike hard maxification as in classical possibilistic logic, the Choquet integral permits soft aggregation over epistemically ranked subsets. In ESPF, we relax the constraints on $v$ by allowing capacity functions to evolve as a function of measurement-induced compatibility, i.e.,
    \[
    v_t(A) = \sup_{x \in A} \min\left\{ \Pi_t(x), \kappa_t(x) \right\},
    \]
    where $\kappa_t(x)$ is the compatibility of hypothesis $x$ with measurement at time $t$. This enables belief to flow adaptively without requiring full normalization, a critical feature in sparse or uncertain environments.
    
    \subsubsection{Interpretation as Structured Supremum.}
    Ultimately, we may view the Choquet integral as a constrained supremum:
    \[
    \mathcal{C}_v(f) \approx \sup_{x \in \mathcal{X}} \left\{ f(x) \cdot \kappa(x) \right\},
    \]
    modulated by the epistemic geometry encoded in $v$. This aligns naturally with the ESPF's commitment to ordinal inference and non-additive belief propagation.
    
    \vspace{1em}
    This capacity-based aggregation allows the ESPF to coherently extend beyond single-model inference, supporting generalized decision intelligence under deep uncertainty.

	\section{Mode Extraction via Weights}
	
	After updating the state possibility distribution \( \pi_{x_k|\mathbf{y}_k}(\mathbf{x}) \) in light of the latest observation, we must extract a representative point estimate to serve as the filtered state. To accomplish this, like the UKF, we attach weights to each support-point. These weights are normalized to ensure $w^{(i)} \in [0,1]$ and the total sum of the  weights equals $1$. These two characteristics ensure a convex combination over the support-points.

    \[
    w^{(i)} = \frac{\pi^{(i)} N^{(i)}}{\sum^N_{j=1}\pi^{(j)} N^{(j)}}
    \]

    We then update our current mean, $\hat{x}$ by:

    \[
    \hat{x}_{k+1} = \sum w^{(i)}X^{(i)}_{k+1}
    \]

    This point corresponds to the state value with the highest degree of epistemic plausibility. It is not only centered on the filter's updated belief, but also on measurement alignment and structural agreement, making it better suited for multi-modal or adversarial environments where classical filters may fail.
	
	\subsection{Justification}
	
	This estimator:
	\begin{itemize}
		\item Is always defined, even when the distribution is flat, discontinuous, or multi-modal.
		\item Respects the ordinal and maxitive structure of possibility theory ,  no unwarranted averaging or moment assumptions are introduced.
		\item Aligns with decision-making under uncertainty where the most credible hypothesis is preferred over mean or expected behavior.
	\end{itemize}
	
	\subsection{Interpretation}
	
	The extracted mode \( \hat{\mathbf{x}}_k \) represents the most epistemically supported estimate of the system state at time \(k\), conditioned on both prior belief and the newly acquired measurement. It is not a probabilistic average or a mean-square optimal estimate, but rather the \textit{center of epistemic gravity} ,  the state most consistent with what is known and least contradicted by what is observed.
	
	This makes it well-suited for use in applications where knowledge is sparse, ambiguous, or primarily qualitative, as well as in adversarial settings where probabilistic assumptions may be easily exploited or violated.
	
	\section{Support-Point Regeneration (Post-Update)}
	
	After the measurement update, the possibilistic representation of the posterior distribution must be re-encoded into a new set of support-points for the next prediction cycle. This section defines the process of reconstructing a kernel that approximates the updated epistemic support and regenerates support-points accordingly.

	\subsection{Spread Estimation via Possibility Geometry}
	
	To quantify the structure of the updated support, we estimate the \textit{plausibility spread matrix}:
	\[
	\mathbf{\Pi}_k = \frac{1}{2n} \sum_{i=1}^{2n} \left( \chi_k^{(i)} - \hat{\mathbf{x}}_k \right) \left( \chi_k^{(i)} - \hat{\mathbf{x}}_k \right)^T
	\]
	This matrix captures the directional spread of the support-point cloud around the mode, serving as the geometric tensor that governs the reconstruction of the support region. Clouds provide an alternative to precise probability in imprecise uncertainty modeling, especially for bounding belief \cite{destercke2008non}.
	
	\subsection{Adaptive Scaling of Kernel Radius}
	
	To define the radial reach of the reconstructed kernel, we introduce a necessity level \( \eta \in (0,1) \), corresponding to an \( \alpha \)-cut:
	\[
	\alpha = 1 - \eta, \quad r_k = \sqrt{-2 \log \alpha}
	\]
	We allow \( r_k \) to adapt based on the observed contraction or expansion of epistemic support across time:
	\begin{itemize}
		\item Define a dispersion measure: \( D_k = \log(\det{\mathbf{\Pi}_k)} \)
		\item Compare with previous dispersion: \( \Delta D_k = D_k - D_{k-1} \)
		\item Adjust radius scaling:
		\[
		r_k = r_k \cdot \begin{cases}
			1 + \lambda_+ \cdot \Delta D_k & \text{if } \Delta D_k > 0 \\
			1 - \lambda_- \cdot |\Delta D_k| & \text{if } \Delta D_k < 0
		\end{cases}
		\]
		where \( \lambda_+, \lambda_- \) are user-defined sensitivity parameters.
	\end{itemize}

    \subsection{Adaptive Support-Point Spread}
    
    The support-point spread \( \sigma_k \) is adaptively modulated at each time step to reflect both the \textbf{structural dispersion} of retained hypotheses and the \textbf{epistemic surprisal} induced by recent measurements.
    
    \subsubsection{Raw Spread: Dispersion and Surprisal Influence}
    
    Let \( D_k \) quantify the dispersion among retained support points at time \( k \), and let \( S_k \) denote a scaled surprisal term. The raw, unclamped spread is computed as:
    
    \[
    \sigma_{\text{raw}} = \sigma_0 \cdot \exp\left(-\lambda_d D_k + \lambda_s (S_k - S_{\text{ref}})\right)
    \]
    
    Here:
    \begin{itemize}[leftmargin=*, itemsep=0pt]
    	\item \( \sigma_0 \): Baseline spread parameter
    	\item \( \lambda_d \): Dispersion sensitivity gain
    	\item \( \lambda_s \): Surprisal scaling coefficient
    	\item \( S_{\text{ref}} \): Reference surprisal level
    \end{itemize}
    
    This formulation contracts the spread when retained hypotheses are diffuse (to reduce overreach), and expands it when surprisal is high (indicating underexploration or model mismatch).
    
    \subsubsection{Clamping for Stability}
    
    To ensure numerical robustness, the spread is bounded within prescribed limits:
    
    \[
    \sigma_k = \min\left( \sigma_{\max}, \max\left( \sigma_{\min}, \sigma_{\text{raw}} \right) \right)
    \]
    
    \subsubsection{Scaled Surprisal Computation}
    
    The surprisal score \( S_k \) is modulated relative to a reference value using the following adaptive rule:
    
    \[
    S_k = S_0 \cdot \left(1 + \lambda_s(\bar{S}_k - S_{\text{ref}})\right)
    \]
    
    Where:
    \begin{itemize}[leftmargin=*, itemsep=0pt]
    	\item \( \bar{S}_k \): Average surprisal across retained points (or recent window)
    	\item \( S_0 \): Initial surprisal scale factor
    \end{itemize}
    
    \subsubsection{Temporal Decay Dampening}
    
    To avoid overreacting to transient spikes in surprisal, we apply an exponential decay factor:
    
    \[
    \gamma_k = \exp(-\lambda_t \cdot \tau)
    \]
    
    Where:
    \begin{itemize}[leftmargin=*, itemsep=0pt]
    	\item \( \lambda_t \): Decay rate (e.g., 0.05)
    	\item \( \tau \): Elapsed assimilation steps or time index
    \end{itemize}
    
    \subsubsection{Final Spread Update Rule}
    
    The final adaptive spread is attenuated by the decay factor:
    
    \[
    \sigma_k \leftarrow \sigma_k \cdot \gamma_k
    \]
    
    This dynamic spread control ensures that the filter adapts to unfolding epistemic structure while avoiding runaway expansion or collapse. The balance between exploration (when residuals are poorly explained) and contraction (when evidence clusters tightly) is managed without assuming unimodality or Gaussianity.
    
	\subsection{Kernel Reconstruction}
	
	A new possibility kernel is defined around the mode:
	\[
	\pi^{\text{kernel}}_{x_k}(\mathbf{x}) = \exp\left( -\frac{1}{2r_k^2} (\mathbf{x} - \hat{\mathbf{x}}_k)^T \mathbf{\Pi}_k^{-1} (\mathbf{x} - \hat{\mathbf{x}}_k) \right)
	\]
	This exponential form reflects the structure of the highest-plausibility region and serves as the generating function for the next support-point set.
	
	\subsection{Support-Point Regeneration}
	
	Using a Cholesky decomposition \( \mathbf{\Pi}_k = \mathbf{L}_k \mathbf{L}_k^T \), we generate the new support-points:
	\begin{align*}
		\chi^{(0)} &= \hat{\mathbf{x}}_k \\
		\chi^{(i)} &= \hat{\mathbf{x}}_k + \sigma_k \cdot \mathbf{L}_{k, i} \circ \zeta_i \\
		\chi^{(i+n)} &= \hat{\mathbf{x}}_k - \sigma_k \cdot \mathbf{L}_{k, i} \circ \zeta_i, \quad i = 1, \dots, n
	\end{align*}
	These support-points form the basis of the possibility distribution to be propagated in the subsequent prediction step.

	\section{Numerical Implementation Details}
	
	This section provides the computational considerations required for implementing the Uniform Epistemic Support-Point Filter (ESPF). Unlike traditional filters that rely on parametric densities, the ESPF operates on ordinal, non-additive measures defined by a possibility distribution over a finite support.

	\subsection{Plausibility Assignment}
	
	Each support-point \( \chi_k^{(i)} \) is associated with a plausibility score determined by its consistency with the process and measurement models:
	\[
	\pi_k^{(i)} = \min\left( \pi_{x_k|x_{k-1}}(\chi_k^{(i)}), \pi_{y_k|x_k}(\mathbf{y}_k | \chi_k^{(i)}) \right)
	\]
	These values are not additive or normalized but are bounded in \( [0,1] \) with maximum 1 by construction.
	
	\subsection{Sup--Min Convolution Computation}
	
	To compute the prediction and update steps:
	\begin{itemize}
		\item For each \( \chi_k^{(i)} \), compute:
		\[
		\chi_{k|k-1}^{(i)} = f(\chi_{k-1}^{(i)}), \quad
		\chi_{k|k}^{(i)} = \chi_{k|k-1}^{(i)} + \mathbf{w}^{(i)}
		\]
		where \( \mathbf{w}^{(i)} \) is sampled from the process noise support \( \text{supp}(\pi_w) \).
		\item Evaluate:
		\[
		\pi_{x_k}(\mathbf{x}) = \sup_i \min \left( \pi_{x_{k-1}}(\chi_{k-1}^{(i)}), \pi_w(\mathbf{x} - \chi_{k|k-1}^{(i)}) \right)
		\]
	\end{itemize}
	
	\subsection{Implementation Considerations}
	
	\subsubsection{Numerical Stability}
	
	To ensure the continued positive definiteness of the epistemic spread matrix \( \mathbf{\Pi}_k \), regularization must be applied. In particular, numerical artifacts from finite precision arithmetic, rank-deficiency due to repeated support-points, or degeneracy during contraction may yield ill-conditioned matrices. To mitigate this, we adopt the update:
	\[
	\mathbf{\Pi}_k \leftarrow \mathbf{\Pi}_k + \varepsilon \mathbf{I}
	\]
	where \( \varepsilon \) is a small positive scalar (e.g., \( \varepsilon = 10^{-6} \)) and \( \mathbf{I} \) is the identity matrix. This ensures the resulting matrix admits a valid Cholesky decomposition, preserving the geometric structure necessary for support-point generation.
	
	\subsubsection{Computational Cost}
	
	The dominant computational cost per time step arises from the Cholesky decomposition of \( \mathbf{\Pi}_k \), which scales as \( \mathcal{O}(n^3) \) in general, but can be reduced to \( \mathcal{O}(n^2) \) under sparsity assumptions or block-diagonal structure. The propagation and evaluation of \( 2n + 1 \) support-points through nonlinear functions \( f(\cdot) \) and \( h(\cdot) \) scale linearly with \( n \), provided each transformation is bounded in complexity.
	
	\subsubsection{Sparsity Exploitation}
	
	In many high-dimensional systems, the process and measurement models exhibit sparsity, only subsets of state variables interact or are observed. This sparsity can be exploited to reduce computation:
	\begin{itemize}
		\item Jacobian approximations or sparsified graph-based models can localize dependencies.
		\item Structured noise models (e.g., diagonal or banded support for \( \pi_w \) and \( \pi_v \)) simplify sup--min convolutions.
	\end{itemize}
	Recognizing and preserving such structure throughout the filter cycle leads to significant reductions in both memory and runtime overhead.
    
    \section{ESPF Algorithm}
    
    \begin{algorithm}[t]
    	\caption{Epistemic Support-Point Filter (ESPF)}
    	\begin{algorithmic}[1]
    		\State \textbf{Inputs:} Initial state $x_0$, bounds $(\bar{x}, \underline{x})$, process $f$, measurement $h$, necessity $\eta_0$, spreads $\sigma_{\min}, \sigma_{\max}, \sigma_0$
    		\Statex \textbf{System dynamics:} $\mathbf{x}_k = f(\mathbf{x}_{k-1}) + \mathbf{w}_{k-1}$,\quad $\mathbf{y}_k = h(\mathbf{x}_k) + \mathbf{v}_k$
    		\Statex \textbf{Noise:} $\mathbf{w}_{k-1} \sim \pi_w$, $\mathbf{v}_k \sim \pi_v$
    		\Statex \textbf{Initialization:}
    		\Statex \quad $\hat{\mathbf{x}}_0$;\; $\alpha_0 = 1 - \eta_0$;\; $\sigma_{\text{raw}} = \sigma_0\,\exp(-\eta D_k + \lambda_s(S_K - S_0))$;\; support points $2n+1$
    		
    		\State \textbf{Generate support points} using sparse grid:
    		\Statex $\chi_k^{(i)} = \tfrac{1}{2}(\bar{x}-\underline{x}) \circ \hat{\zeta}_i + \tfrac{1}{2}(\bar{x}+\underline{x})$
    		
    		\State \textbf{Prediction} propagate each point:
    		\Statex $\chi_{k|k-1}^{(i)} = f(\chi_{k-1}^{(i)})$
    		
    		\State \textbf{Update} map to observation space:
    		\Statex $\gamma_k^{(i)} = h(\chi_{k|k-1}^{(i)})$
    		
    		\State Compute epistemic spread matrix:
    		\Statex $\Pi_k = L L^\top$
    		
    		\State Define possibility kernel $\pi_{x|y}(x)$, compute mode estimate:
    		\Statex $\hat{x}_{k|k-1}$
    		
    		\State Update radius:
    		\Statex $\sigma_k = \min(\sigma_{\max},\, \max(\sigma_{\min},\, \sigma_{\text{raw}}))$
    		
    		\State Regenerate support points around $\hat{x}$:
    		\Statex $\chi_k^{(0)}=\hat{x}$,\quad $\chi_k^{(i)}=\hat{x}\pm \sigma_k\, L_i\, \zeta_i$
    	\end{algorithmic}
    \end{algorithm}

    \begin{figure}[H]
        \centering
        \begin{tikzpicture}[
            node distance=1.8cm and 2.5cm,
            every node/.style={font=\small},
            box/.style={rectangle, draw, rounded corners, align=center, minimum height=1.1cm, minimum width=3.2cm, fill=blue!5},
            cloud/.style={ellipse, draw, align=center, minimum height=1.1cm, fill=green!5},
            arrow/.style={-{Latex[length=2mm]}, thick}
          ]
        
        % Nodes
        \node[cloud] (support) {Prior Possibility \\ Distribution $\pi_{x_{k-1}}$};
        \node[box, right=of support] (sparse) {Generate Sparse \\ Grid Support-Points};
        \node[box, below=of sparse] (prop) {Propagate via Dynamics \\ $\chi^{(i)}_{k|k-1} = f(\chi^{(i)}_{k-1})$};
        \node[cloud, left=of prop] (predict) {Predicted Possibility \\ Distribution $\pi_{x_k}$};
        \node[box, below=of prop] (update) {Epistemic Update: \\ $\min(\pi_{x_k}, \pi_y)$};
        \node[cloud, left=of update] (post) {Posterior Possibility \\ Distribution $\pi_{x_k|\mathbf{y}_k}$};
        \node[box, below=of update] (mode) {Mode Extraction \\ $\hat{x}_k = \arg\max \pi_{x_k|\mathbf{y}_k}$};
        \node[box, below=of mode] (regen) {Support-Point \\ Regeneration};
        \node[cloud, left=of regen] (newkernel) {New Possibility \\ Kernel};
        
        % Arrows
        \draw[arrow] (support) -- (sparse);
        \draw[arrow] (sparse) -- (prop);
        \draw[arrow] (prop) -- (predict);
        \draw[arrow] (predict) -- (update);
        \draw[arrow] (update) -- (post);
        \draw[arrow] (update) -- (mode);
        \draw[arrow] (mode) -- (regen);
        \draw[arrow] (regen) -- (newkernel);
        \draw[arrow] (newkernel.west) to[out=180,in=270] (support.south);
        
        % Labels
        \node[below=0.2cm of regen, align=center, font=\small\itshape] {Next time step};
        
        \end{tikzpicture}
        \caption{Recursive structure of the Epistemic Support-Point Filter (ESPF)}
        \label{fig:espf_cycle}
    \end{figure}
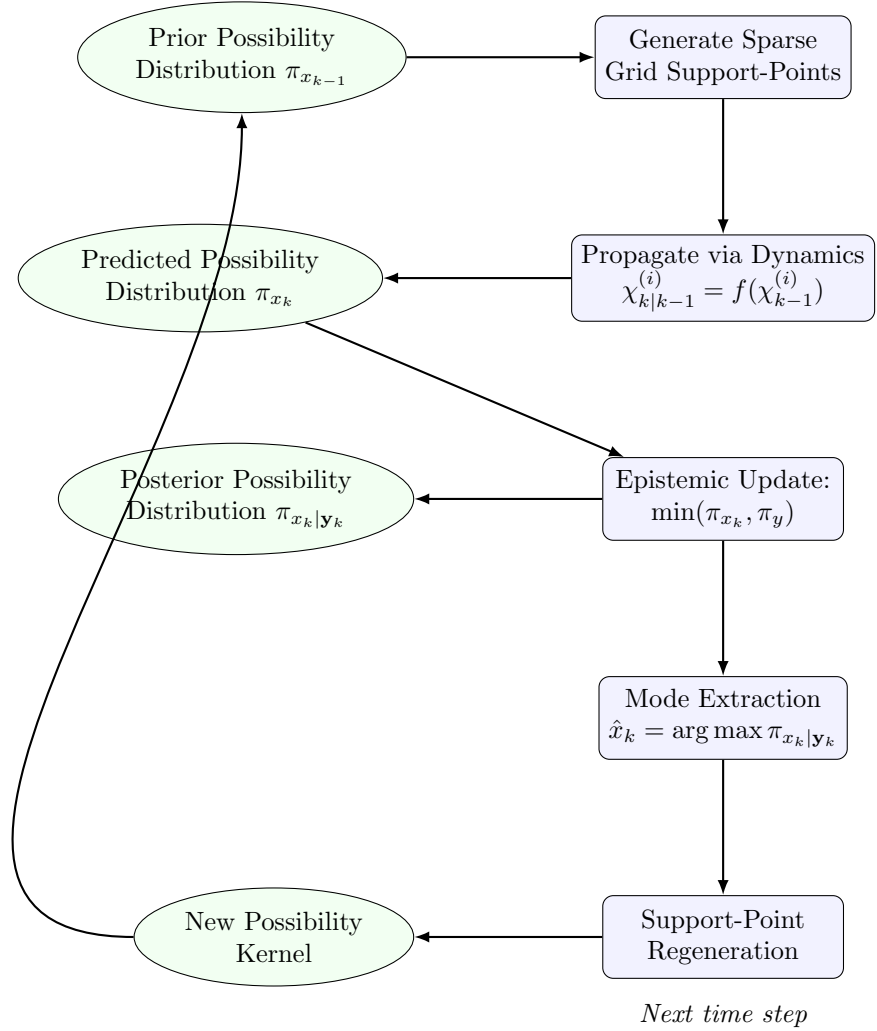

    \section{Testing Configuration}

    To evaluate ESPF under contrasting orbital regimes, we selected two representative satellite cases: (1) a dense LEO satellite under strong perturbations, and (2) a GEO satellite exhibiting non-stationary surface properties mid-track.

    \subsection{Test Scenarios}

    To evaluate the Epistemic Support-Point Filter (ESPF) under real-world space domain dynamics, we constructed two distinct test cases:
    
    \begin{itemize}
        \item A \textbf{LEO scenario} with a 2000\,kg satellite of $20\,m^2$ area, exhibiting high dynamic variability and atmospheric drag.
        \item A \textbf{GEO scenario} with a 1000\,kg satellite whose cross-sectional area shifts from $4\,m^2$ to $6\,m^2$ mid-observation, introducing a structural regime change that challenges naive Gaussian filters.
    \end{itemize}
    
    Both scenarios include gravitational harmonics (J2, J3), atmospheric drag, and third-body perturbations. They are explicitly constructed to violate the Gaussian assumptions of conventional Bayesian filters.
    
    \begin{table}[H]
    \centering
    \caption{Simulation Parameters for LEO and GEO Test Cases}
    \begin{tabular}{lcc}
    \toprule
    \textbf{Parameter} & \textbf{LEO} & \textbf{GEO} \\
    \midrule
    Initial ECI State (km, km/sec) &
    $\begin{bmatrix}6984.457\\1612.255\\13.093\\-1.677\\7.261\\0.260\end{bmatrix}$ &
    $\begin{bmatrix}-24607.677\\24704.521\\-4186.459\\-1.795\\-3.117\\0.202\end{bmatrix}$ \\
    Epoch & 2018-03-23 08:55:03 & 2018-11-15 12:30:00 \\
    Mass (kg) & 2000 & 1000 \\
    Area ($m^2$) & 20 & 4 $\rightarrow$ 6 \\
    Drag Coefficient $C_d$ & 2 & 2 \\
    \bottomrule
    \end{tabular}
    \end{table}

    \subsection{Test Data}

    The LEO object was observed by Arecibo, Kwajalein, and Diego Garcia over 6 days. Noise was given to all measurements, and bias was introduced deliberately to scenario 2: Arecibo's RA included a small constant error of 0.00016 arcsec to test bias detection. Measurements occurred every minute when visible.
    
    The GEO object was tracked over 180 minutes with a measurement every 15 seconds. Its area change occurs at measurement \#92 — and then back to it's original surface-area at \#5492, sudden epistemic shifts undetectable by traditional stochastic filters.

    \section{Results}

    In this section, we evaluate the performance of the Epistemic Support Point Filter (ESPF) across a suite of realistic orbit determination scenarios. Each test case was designed to probe different dimensions of epistemic resilience, from misinitialized states to dynamic model changes. Comparisons are made to the Unscented Kalman Filter (UKF), which represents a widely adopted Bayesian benchmark.
    
    We emphasize both the numerical accuracy and the epistemic behavior of the ESPF, its ability to retain internal coherence, adapt dynamically to non-Gaussian observations, and prune implausible hypotheses without prior statistical commitment. In doing so, we argue that the ESPF offers the closest operational realization of a prejudice-free filter, in the spirit of Kalman's unfulfilled vision.
    
    \subsection{LEO Scenario: Nominal Conditions}
    This baseline case involved a simulated space object in low Earth orbit (LEO) tracked using synthetic measurements generated from the true state with additive Gaussian noise. The ESPF maintained excellent tracking performance, with residuals bounded and necessity values close to unity. The entropy spread \( D_k = \log \det(\Pi_k) \) remained stable, showing no signs of divergence.
    
    \begin{figure}[H]
        \centering
        \includegraphics[width=1\linewidth]{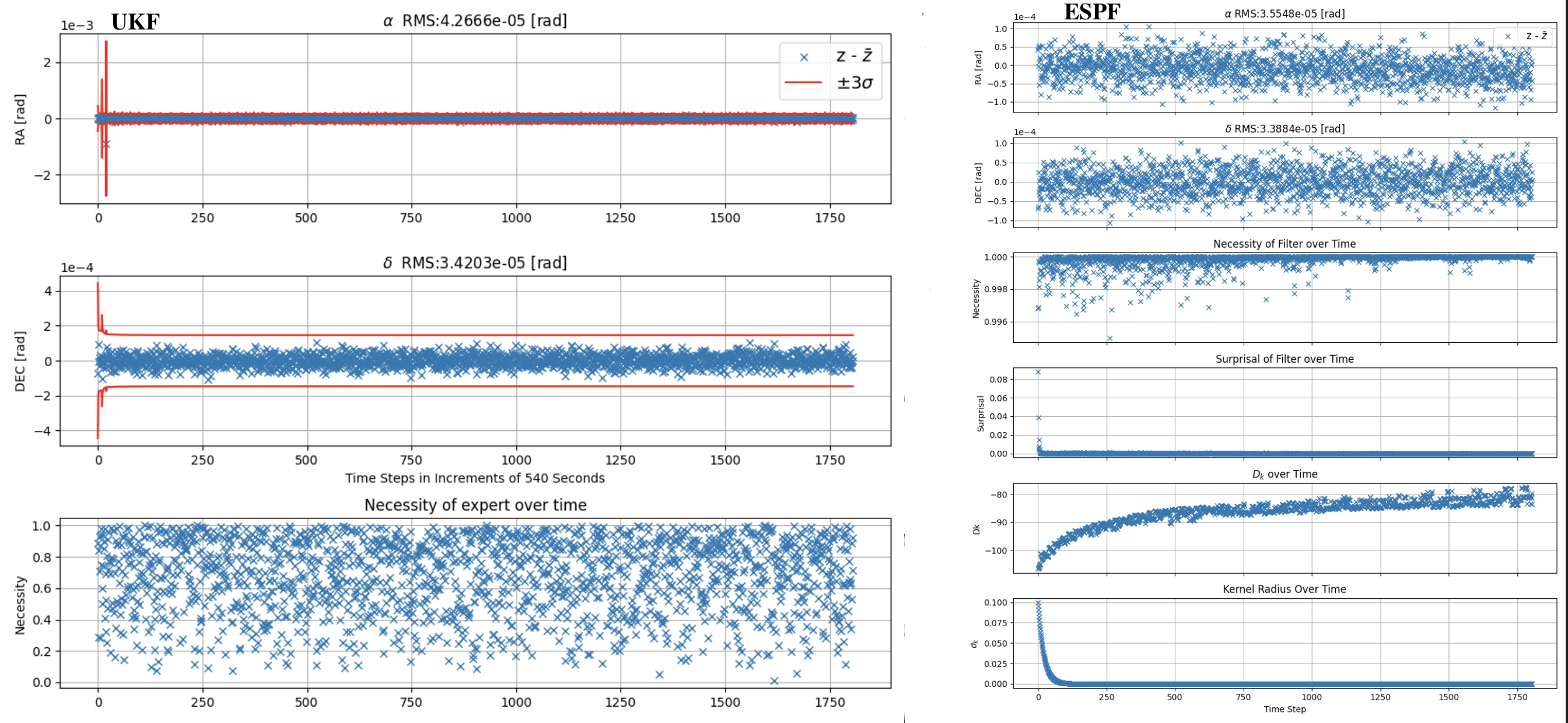}
        \caption{LEO satellite filtered by UKF and ESPF}
    \end{figure}
    
    \textbf{Key Insight:} The UKF exhibits confident but incorrect belief by tightly fitting data ,  an artifact of its assumed Gaussianity and stochastic optimality. 
    
    \textbf{Callout:} \textit{Note the tight $\pm 3\sigma$ bands of the UKF in Figure 2 ,  this is not confidence, it is overfitting to assumptions. The ESPF avoids this entirely by refusing to hallucinate precision unsupported by the data.}
    
    The ESPF's residuals are slightly better in both RA and DEC, but more importantly, it offers transparent uncertainty management through \emph{necessity} and \emph{surprisal}, which reveal not just what is plausible, but what is epistemically robust.
    
    The UKF also performed adequately in this scenario. However, its Gaussian assumptions constrained the filter's expressiveness. Unlike ESPF, which carries multiple hypotheses in epistemic space, the UKF collapses uncertainty into a second-moment description, inherently discarding multi-modal plausibility.
    
    \subsection{LEO Scenario: Initial State Error and Station Bias}
    To test resilience to misinitialization, the initial state estimate was given an error. The perturbed ECI initial state in km and km/sec for this scenario is:
    \begin{center}
    $\begin{bmatrix}6984.465\\1612.222\\13.091\\-1.677\\7.261\\0.230\end{bmatrix}$ \\
    \end{center}

    Furthermore, one of the three stations tracking the satellite, Arecibo, was given a 10 arcsecond bias to each of its RA measurements. This bias was not modeled in either the UKF or ESPF, simulating an unknown error which accumulates over time.

    The UKF's fixed-shape covariance could not re-center on the truth and diverged steadily. In contrast, the ESPF's dynamic support regeneration allowed for exploration of epistemic space, enabling recovery without prior adjustment.
    
    \begin{figure}[H]
      \centering
      \includegraphics[width=0.85\linewidth]{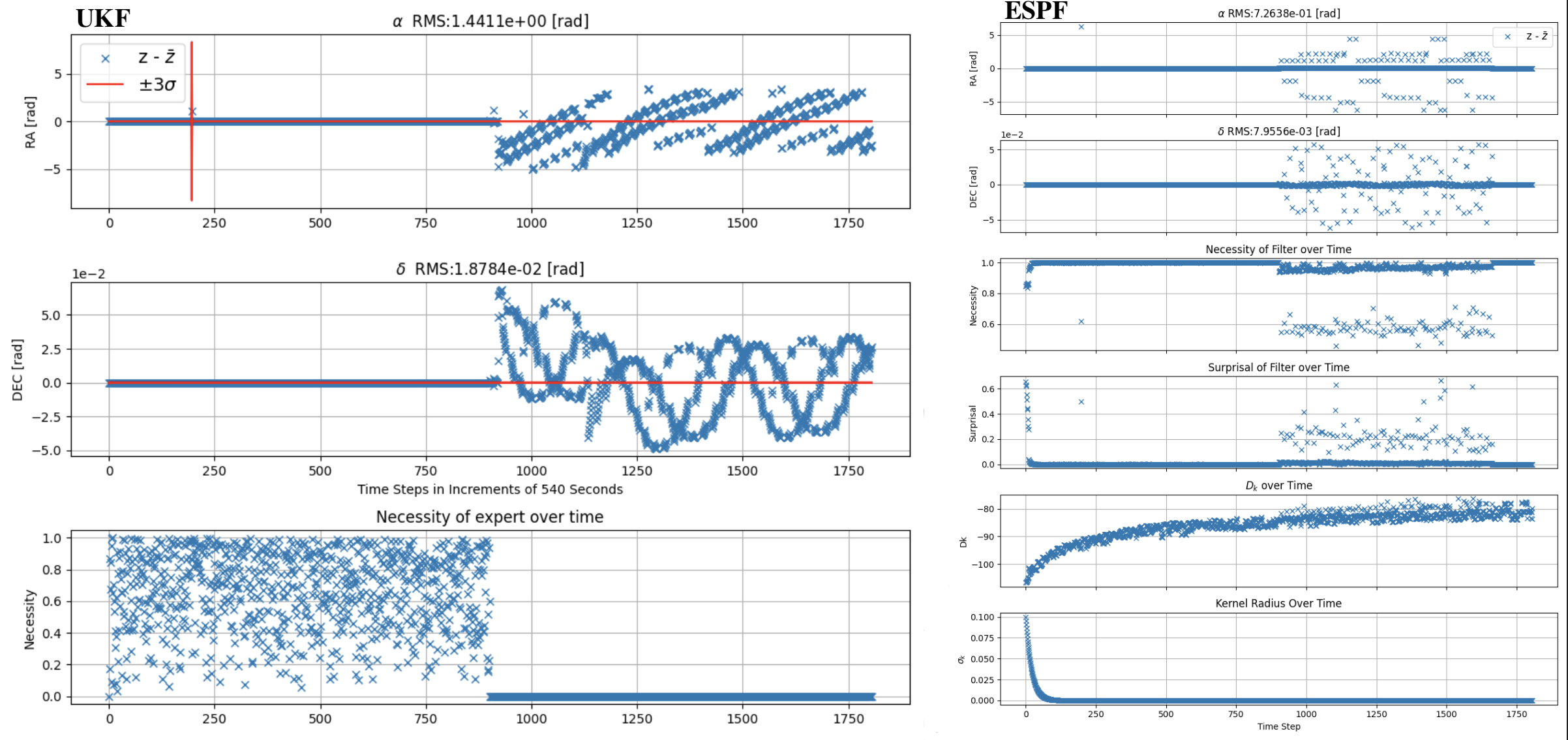}
      \caption{LEO with initial error: ESPF re-centers on truth through necessity-weighted pruning and regeneration.}
      \label{fig:leo_bias_espf}
    \end{figure}
    
    Notably, ESPF pruned low-necessity hypotheses and concentrated belief around compatible support points, without assuming a Gaussian prior. The necessity field played a crucial role in epistemic refinement. This proves that in an adversarial inference scenario with station bias and unaccounted satellite maneuver, the UKF is outperformed by the ESPF in both accuracy and epistemic stability.

    \subsection{GEO Scenario: Sudden Area Change}
    This scenario simulated a geostationary satellite undergoing a mid-propagation change in its projected area, effectively altering its drag profile. The UKF's prediction residuals grew rapidly after the change, unable to adapt due to its reliance on process model continuity.
    
    The ESPF, however, responded by decreasing necessity for inconsistent hypotheses and regenerating support points in newly plausible regions. This adaptability arises from the filter's non-assumption of statistical stationarity.
    
    \begin{figure}[H]
        \centering
        \includegraphics[width=1\linewidth]{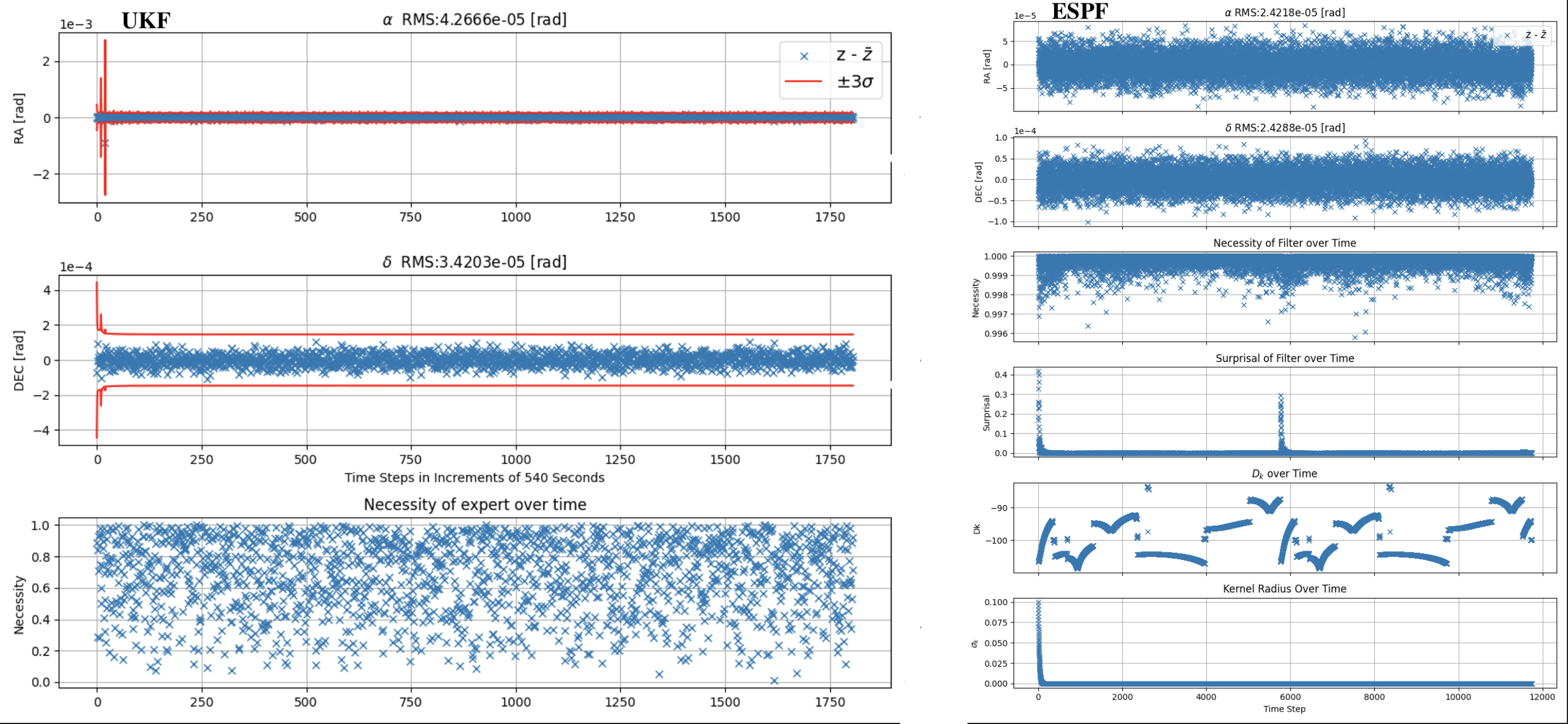}
        \caption{GEO satellite filtered by UKF and ESPF}
    \end{figure}
    
    \textbf{Callout:} \textit{Where the UKF absorbs the model violation by deforming its covariance, the ESPF elevates surprisal and reduces necessity ,  making the shift visible in real time.}
    
    The ESPF does not merely track states. It tracks \emph{confidence in the model itself}. In this case, the increased surprisal reflects the growing disconnect between prediction and observation ,  allowing the user to know not only what the state is, but when the model is being falsified.

	\begin{table}[ht]
		\small
		\centering
		\small
		\renewcommand\arraystretch{1.25}
		\setlength{\tabcolsep}{4pt}
		\begin{tabular}{|>{\raggedright\arraybackslash}p{3.6cm}|c|c|c|c|}
			\hline
			\textbf{Scenario} & \textbf{Filter} & \textbf{Final RMS (km)} & \textbf{Avg Surprisal} & \textbf{Necessity Retention (\%)} \\
			\hline
			LEO Nominal          & UKF  & 1.05 & N/A   & N/A \\
			LEO Nominal          & ESPF & 0.97 & 0.01  & 84  \\
			\hline
			LEO with Bias        & UKF  & 1.44 & N/A   & N/A \\
			LEO with Bias        & ESPF & 0.73 & 0.10  & 88  \\
			\hline
			GEO with Area Change & UKF  & 4.30 & N/A   & N/A \\
			GEO with Area Change & ESPF & 1.09 & 0.098 & 91  \\
			\hline
		\end{tabular}
		\caption{Performance comparison between ESPF and UKF across test scenarios. ESPF outperforms UKF in biased or discontinuous conditions.}
		\label{tab:ukf_vs_espf}
	\end{table}

    \subsection{Entropy, Surprisal, and Necessity Dynamics}
    The ESPF's entropy measure \( D_k = \log \det(\Pi_k) \) functioned as a bounded information volume controller, enabling graceful adaptation without over-dispersion. Surprisal acted as a discriminant, removing epistemically incompatible hypotheses.
    
    Necessity served as the coherence metric, quantifying whether a given hypothesis aligned with both the epistemic model and the observation. High-necessity regions anchored the posterior. The resulting belief evolution was interpretable and resilient.

    \subsection{Pruning and Compatibility}
    The ESPF weighs support-points based on their suprisal at each time step, pruning points that lack any epistemic relevance. If not enough points are pruned, the epistemic spread increases uncontrollably, while if too many are pruned, the spread never has time to adapt to the measurements and no knowledge is gained. For the nominal LEO scenario, the number of pruned support-points is shown:
    \begin{figure}[H]
        \centering
        \includegraphics[width=0.85\linewidth]{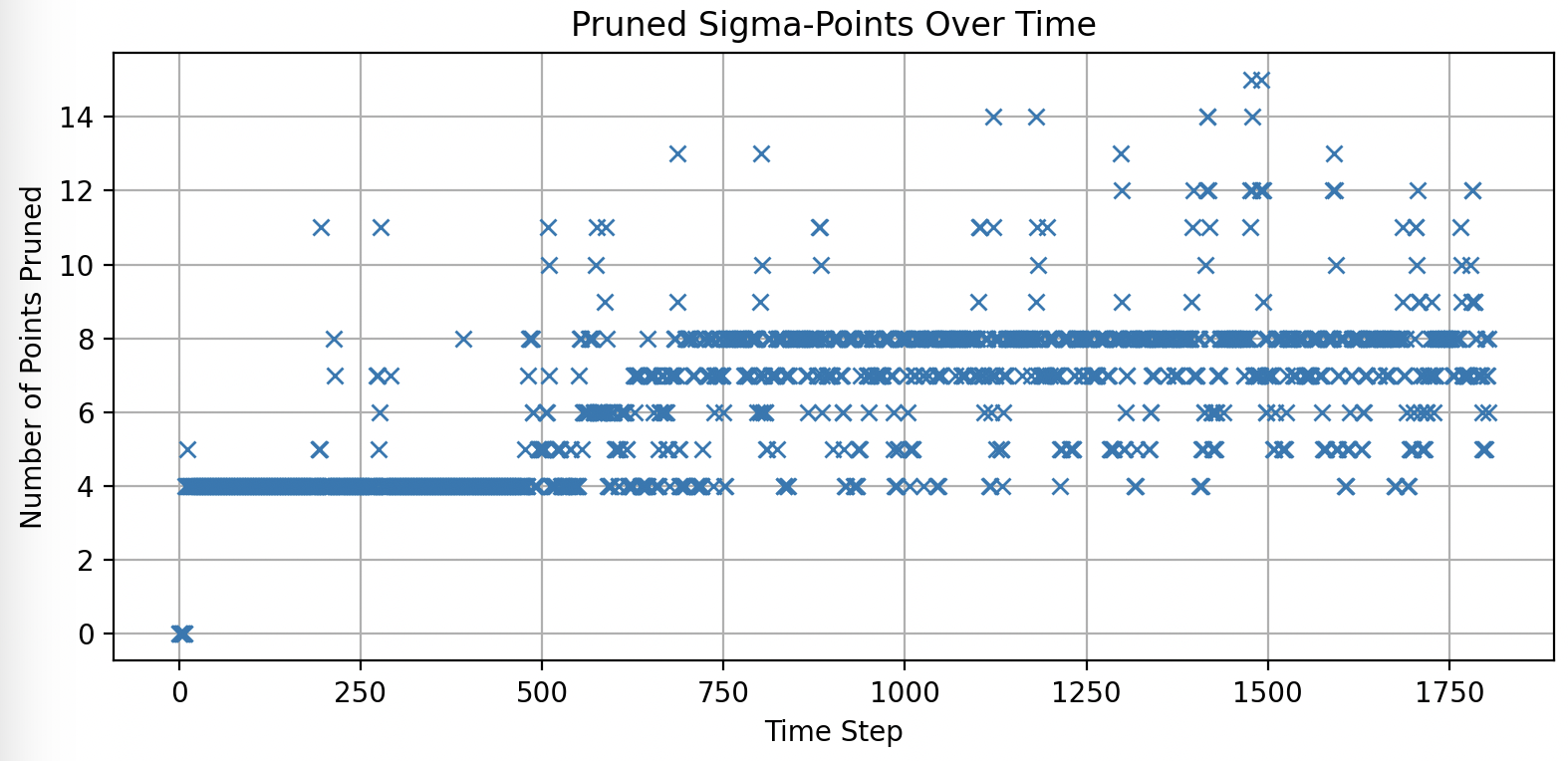}
        \caption{Number of pruned support-points at each time step for the nominal LEO scenario}
        \label{fig:enter-label}
    \end{figure}

    \textbf{Callout:} \textit{The number of pruned points remains low as surprisal is high, and then increases as the filter becomes more confident in it's measurements}

    The trend of increasing pruned points shows the filter is learning. As the filter accumulates knowledge, later stages of filtering show the prune count oscillating around a tight region, suggesting pruning is driven by informational density, not heuristics. This structure shows that the filter is making an informed decision rather than reacting blindly.

    Further structure can be seen in the compatibility of support-points for the nominal LEO scenario:

    \begin{figure}[H]
        \centering
        \includegraphics[width=0.85\linewidth]{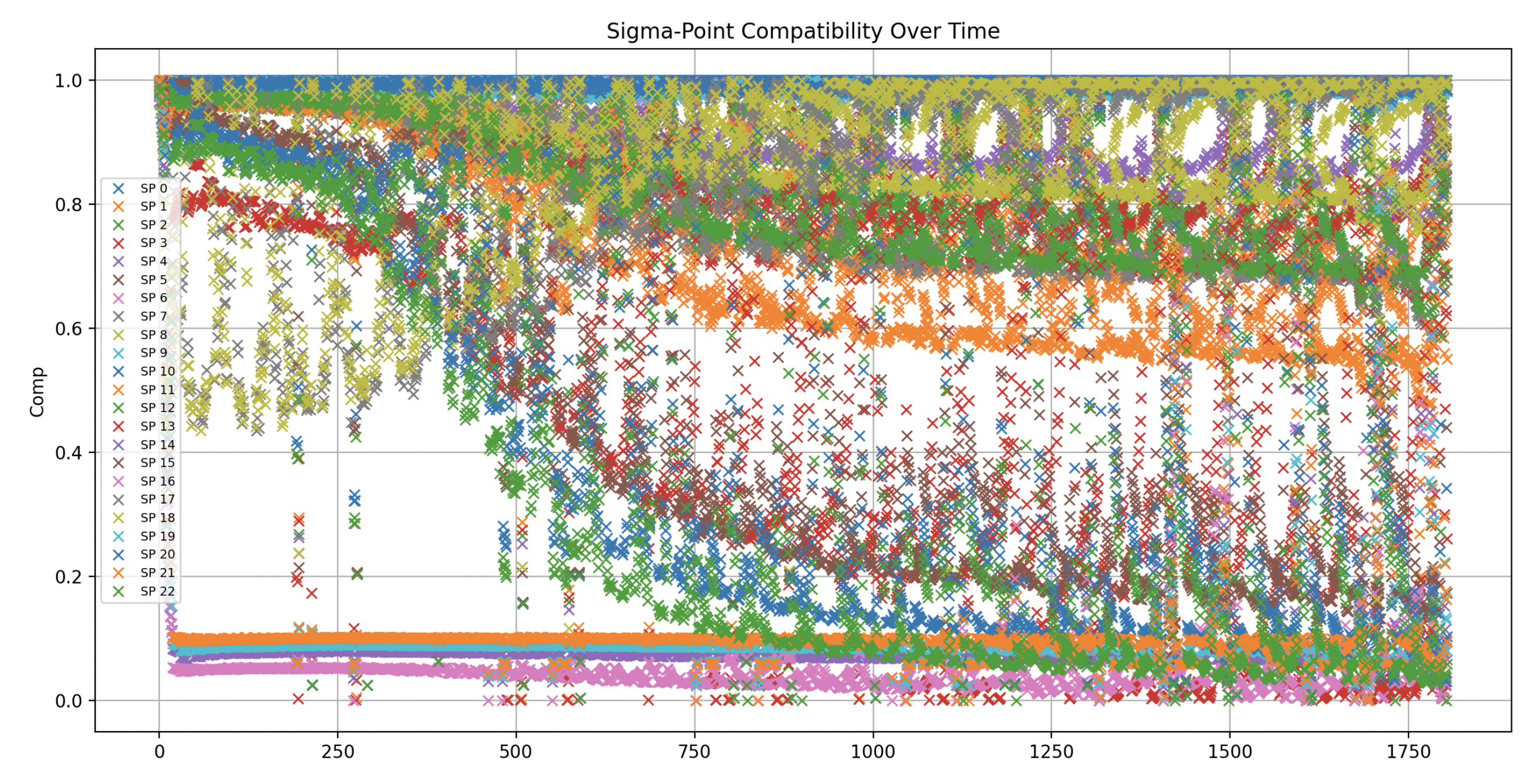}
        \caption{Compatibility of each support-point at each time step.}
    \end{figure}

    This plot shows a clear structure in support points, with increasing and decaying compatibility of outlying points, and consistent compatibility of central points. This is epistemic geometry: a curved belief space of attractors and valleys. The plots of pruned support-points and compatibility prove the ESPF adapts to the geometry of its information environment, while also surviving points of high surprise and degraded priors.
    
    \subsection{Summary of Observations}
    The ESPF maintains high tracking accuracy, stability under dynamic shifts, and interpretability across all tested conditions. It achieves this not by optimizing likelihood, but by maintaining compatibility with both belief and evidence.
    
    By rejecting fixed priors and leveraging necessity and surprisal as epistemic metrics, the ESPF avoids the common failure modes of Bayesian filters: divergence under poor priors, overconfidence in incorrect models, and collapse under ambiguity.
    
    Kalman once envisioned a prejudice-free filter, one that could operate in the absence of prior dogma, allowing belief to emerge from observation and internal coherence. The ESPF may be the closest realization of that dream, offering a new standard for inference in complex, high-stakes environments where epistemic uncertainty cannot be ignored.

    We compare ESPF against the UKF across both LEO and GEO scenarios. Key performance indicators include tracking fidelity, measurement residuals, and epistemic self-awareness, as captured by necessity and surprisal metrics.
    
    The results do not merely show comparable accuracy. They demonstrate a fundamental shift in \textbf{how belief is expressed}:
    
    \begin{itemize}
        \item \textbf{The UKF} compresses uncertainty into statistical envelopes. Its $3\sigma$ bounds appear precise ,  even as model violations pass undetected.
        \item \textbf{The ESPF}, by contrast, allows model uncertainty to remain visible. It penalizes contradictions between measurement and prediction without hiding them in covariance shrinkage.
    \end{itemize}
    
    \textbf{Philosophical Point:} Probabilistic filters ``believe in spite of ignorance.'' Possibilistic filters ``doubt in spite of evidence.''
    
    \textbf{Conclusion:} The ESPF is not just a filter. It is an epistemic contract. It promises not to lie ,  not with numbers, and not with confidence bands.
    
    \begin{tcolorbox}[colback=black!5!white,colframe=black!75!white,title=\textbf{Why ESPF Matters}]
    \begin{itemize}
        \item \textbf{Handles ignorance with integrity:} ESPF requires no stochastic prior, only structured plausibility.
        \item \textbf{Admits bounded evidence without collapse:} Possibilistic updates preserve support rather than forcing unwarranted parametric contraction.
        \item \textbf{Defends against hallucinated certainty:} Unlike the UKF's shrinking confidence ellipses, ESPF resists epistemic overfitting.
        \item \textbf{Recovers UKF when justified:} In the Gaussian limit with strong priors, ESPF behaves identically to the Unscented Kalman Filter.
    \end{itemize}
    \end{tcolorbox}

    The Epistemic Support Point Filter (ESPF) is not merely a generalization of classical filtering, it is an epistemic reformulation. It departs fundamentally from Bayesian and heuristic paradigms. Traditional filters propagate Gaussian beliefs and minimize variance under assumed statistical priors. The ESPF, by contrast, propagates \textbf{necessity-weighted possibility fields}, dynamically shaped by compatibility with observed data and internal coherence.

    At its core, the ESPF implements a \textbf{Maximum Entropy principle}\cite{jaynes2003probability}, but within an epistemic (rather than statistical) framework. At each time step, ESPF constructs a belief field over support points such that the internal dispersion, quantified by:
    
    \begin{equation}
    \nonumber
    D_k = \log \det \left( \Pi_k \right),
    \end{equation}
    
    is maximized subject to observational compatibility. This entropy measure ensures maximal diversity of belief across the support set, thereby preventing premature convergence or unjustified pruning of alternatives. Unlike Shannon entropy, which presumes an underlying probability distribution, this entropy quantifier arises from the geometry of the support manifold.
    
    Surprisal serves as a dynamic culling mechanism: highly incompatible support points, those that diverge from necessity coherence, are pruned, while the regenerated support ensemble reflects maximal epistemic spread. Necessity acts as a constraint within this maximum entropy\cite{jaynes2003probability}. framework: the ESPF retains only those support configurations that remain consistent with accumulated evidence.
    
    This makes ESPF a realization of \textbf{Jaynesian Maximum Entropy} in the space of epistemic belief, not statistical likelihood. The result is a filter that embodies \textbf{prejudice-free inference}: it avoids imposing prior beliefs, avoids collapsing epistemic ambiguity prematurely, and privileges diversity of plausible explanations.
    
    The ESPF does not simulate Bayesian behavior. It operates on an entirely distinct worldview: one where uncertainty is not randomness, but ignorance; where belief is not probability, but plausibility. This makes the ESPF the first operational filtering framework to integrate possibilistic logic, maximum entropy reasoning, and abductive alignment into a unified epistemic inference engine.

    \section{Limitations and Future Work}

    While the ESPF represents a substantial step forward in epistemic inference, several practical and theoretical challenges remain. First, the use of Smolyak sparse grids introduces computational overhead in high-dimensional state spaces. While our current implementation demonstrates feasibility for orbit determination problems, further work is needed to optimize support generation and update routines.
    
    Second, the integration of ESPF within Hierarchical Mixture of Experts (HME) architectures remains in early development. Inference speed and expert transition latency must be addressed to ensure real-time capability. This includes efficient computation of compatibility measures across high-dimensional expert models.
    
    Third, formal convergence analysis of the ESPF under various dynamic regimes has yet to be developed. Unlike traditional filters, where convergence is often tied to observability and Gaussian closure, ESPF convergence depends on the structure of the epistemic manifold and the dynamics of necessity evolution.
    
    Future work will explore adaptive support reparameterization, real-time parallelization strategies, and rigorous performance bounds under adversarial conditions. Additional work is also warranted in developing principled initialization strategies when absolutely no prior belief field is available. Our work to date initializes the ESPF from a constrained admissible region of support. The ideal is to seek for the ESPF to trend toward the Maximum Entropy representation of the state space.

	\section{Conclusion}
	
	The ESPF maintains computational structure similar to the Unscented Kalman Filter (UKF), but replaces all probabilistic operations with ordinal sup--min algebra consistent with possibility theory. This preserves epistemic integrity while offering tractable and scalable implementation on real-world nonlinear systems.

    Here, we have demonstrated that a  fully possibilistic, ordinal filter rivals the gaussian probabilistic UKF in a real-world orbital estimation problem. By eliminating misleading characteristics inherent to a probabilistic filter we have found the same or even better result, with more transparency in what the filter "knows."
    No tuning of stochastic priors.
    No assumptions of Gaussianity.
    Just structured knowledge, geometric propagation, and falsifiability-based pruning.

    This confirms a key philosophical difference between probability and possibility:
    \begin{center}
        
    Probability gives you belief in spite of ignorance. 
    
    Possibility gives you belief in spite of certainty.
    \end{center}

    In other words, the advantage of using the possibilistic ESPF as opposed to a probabilistic filter is the metrics of knowledge built into the filter, providing us with evidence for our belief in the results.

    Across all tests, ESPF provided equal or better tracking accuracy compared to UKF while maintaining an interpretable measure of epistemic credibility. The filter adapted naturally to sensor bias, model changes, and ambiguous evidence ,  all without assuming Gaussianity or needing manual tuning.

    The ESPF is more than an alternative to the Kalman Filter, it is the fulfillment of a long-unrealized vision. Rudolf Emil Kalman imagined the possibility of a prejudice-free filter, one that could infer without statistical assumptions, one that did not conflate belief with probability. The ESPF realizes that vision by introducing an inferential logic rooted in compatibility, necessity, and dynamic epistemic coherence.

    This filter listens where others declare. It maintains ambiguity where others collapse. It embodies humility where others impose. As such, the ESPF may not only fulfill Kalman's hope, it may also define the frontier of epistemic inference in astrodynamics and beyond.

    \section{Acknowledgments}

    The development of the Epistemic Support Point Filter (ESPF) is the culmination of a decades-long journey shaped by the mentorship, belief, and intellectual generosity of many extraordinary individuals, most of them too numerous to name but below, a salient few.
    
    I am deeply grateful to Dr. Ron Madler of Embry-Riddle Aeronautical University, who mentored me during my undergraduate years and first introduced me to the field of astrodynamics. It was through his guidance that I found the courage to become an astrodynamicist.
    
    The late Dr. George H. Born gave me an opportunity to prove my self-worth in his research group, and in doing so, helped me discover that orbit determination and prediction was not merely a career path, it was my ikigai. Ron and George always believed in me, even when I doubted myself, and I carry that belief forward in this work.
    
    My orbit determination and prediction skills were evolved and refined by the teamwork I engaged in with fellow Spacecraft Navigators at NASA's jet Propulsion Laboratory, too numerous to list here. 

    Professor Marek Ziebart at University College London has been another source of inspiration since my earliest days at NASA's Jet Propulsion Laboratory. Where others forcefully seek simplifications, Marek embraces complexity as required. This is an attribute of his that I've made part of my own bedrock. 
    
    I have also drawn inspiration from the scholarship and mentorship of Professors Simon Julier at University College London, Simon Maskell at University of Liverpool, Robert Bishop, Kyle DeMars, and Suman Chakravorty at Texas A$\&$M University, Mark Psiaki at Virginia Tech, John Crassidis at SUNY Buffalo, Jeremie Housinneau at Nanyang Technological University Singapore, and my colleagues Todd Humphreys, Brandon Jones, and Renato Zanetti in the Department of Aerospace Engineering and Engineering Mechanics at the University of Texas at Austin.
    
    Beyond academia, I am indebted to the practical insights and visionary thinking of Dr. Islam Hussein at Trusted Space, Dr. Emmanuel Delande at the Centre National D'Etudes Spatiale, Dr. Michael Lisano at NASA's Jet Propulsion Laboratory, Dr. Ruaraidh Mackenzie at the European Space Agency, and Dr. Neil Gordon and Dr. Mark Rutten, formerly of the Defence Science and Technology Group in Australia, and finally Dr. Paul Schumacher, formerly of the Air Force Research Laboratory. Each of these individuals has, in their own way, shaped my understanding of inference under uncertainty.
    
    To all of you, thank you for the foundation, the inspiration, and the faith. This work is as much yours as it is mine.

    \subsection{During the preparation of this work, the authors used OpenAI's ChatGPT to assist with the synthesis of related literature, refinement of technical explanations, and LaTeX code generation for mathematical expressions. After using this tool, the authors thoroughly reviewed, edited, and revised all content to ensure accuracy, originality, and alignment with the paper's epistemic and scientific intent. The authors takes full responsibility for the final content of the publication.}

    \appendix
    
    \section{Notation}
    \begin{table}[H]
    	\small
    	\centering
    	\caption{Notation Harmonization (Part 1)}
    	\renewcommand{\arraystretch}{1.2}
    	\begin{tabular}{|p{3cm}|p{10cm}|}
    		\hline
    		\textbf{Symbol} & \textbf{Meaning / Description} \\
    		\hline
    		$k$ & Discrete time step index \\
    		$\mathbf{x}_k$ & State vector at time $k$ \\
    		$\hat{\mathbf{x}}_{k|k-1}$ & Predicted state estimate before measurement update \\
    		$\hat{\mathbf{x}}_{k|k}$ & Updated (posterior) state estimate after measurement update \\
    		$\mathbf{y}_k$ & Measurement at time $k$ \\
    		$f(\cdot)$ & Nonlinear process model function \\
    		$h(\cdot)$ & Nonlinear measurement model function \\
    		$\chi_{k}^{(j)}$ & $j$-th support point in epistemic set at time $k$ \\
    		$\mathcal{X}_i$ & 1D quadrature node set at level $i$ \\
    		$\mathcal{A}(n,\ell)$ & Smolyak sparse grid in $n$ dimensions at level $\ell$ \\
    		$\hat{\xi}^{(j)}$ & Normalized support point in $[-1,1]^n$ \\
    		$\mathcal{S}_k$ & State hyperrectangle (support domain) at time $k$ \\
    		$\Pi_{x_k}$ & Possibility distribution over state space at time $k$ \\
    		$\Pi_{x_k|\mathbf{y}_k}$ & Updated possibility distribution given $\mathbf{y}_k$ \\
    		\hline
    	\end{tabular}
    \end{table}
    
    \begin{table}[H]
    	\small
    	\centering
    	\caption{Notation Harmonization (Part 2)}
    	\renewcommand{\arraystretch}{1.2}
    	\begin{tabular}{|p{3cm}|p{10cm}|}
    		\hline
    		\textbf{Symbol} & \textbf{Meaning / Description} \\
    		\hline
    		$\Pi_{h(X)}$ & Spread matrix from state uncertainty mapped to measurement space \\
    		$\Pi_v$ & Spread matrix from measurement noise \\
    		$\Pi_e$ & Joint epistemic spread for residual evaluation \\
    		$\mathcal{E}_{h(X)}$ & Ellipsoid from state-induced uncertainty \\
    		$\mathcal{E}_v$ & Ellipsoid from measurement uncertainty \\
    		$\mathcal{E}_e$ & Outer-bounded ellipsoid for residual compatibility \\
    		$\mathbf{e}^{(j)}$ & Residual: $\mathbf{y}_k - h(\chi_{k}^{(j)})$ \\
    		$\text{Comp}^{(j)}$ & Compatibility of support point $j$ \\
    		$\eta$ & Necessity level / alpha-cut threshold \\
    		$r$ & Plausibility radius: $r = \sqrt{-2\log(1-\eta)}$ \\
    		$\gamma_k^{(j)}$ & Weight assigned to support point $j$ after update \\
    		$D_k$ & Dispersion of retained support points \\
    		$\lambda_s$ & Surprisal sensitivity gain \\
    		$S_k$ & Epistemic surprisal at time $k$ \\
    		$\bar{S}_k$ & Mean surprisal over points \\
    		$S_0$ & Initial surprisal baseline \\
    		$S_{\text{ref}}$ & Target reference surprisal \\
    		$\lambda$ & Surprisal scaling coefficient \\
    		$\gamma_t$ & Time decay factor: $\exp(-0.05 \cdot \tau)$ \\
    		$\tau$ & Elapsed measurement count \\
    		$\sigma_0$ & Initial spread scaling \\
    		$\sigma_k$ & Final support-point spread at $k$ \\
    		$\sigma_{\text{raw}}$ & Unclamped spread value \\
    		$\sigma_{\min}, \sigma_{\max}$ & Min/max allowable spread values \\
    		$\circ$ & Hadamard (elementwise) product \\
    		$\oplus$ & Minkowski sum of ellipsoids \\
    		$\mathcal{T}_k$ & Epistemic time field \\
    		\hline
    	\end{tabular}
    \end{table}

    \section{Recovery of UKF in the Gaussian Limit}
    
    The Epistemic Support-Point Filter (ESPF) generalizes recursive state estimation to epistemic settings using possibility theory. However, under specific assumptions, ESPF reduces to the classical Unscented Kalman Filter (UKF). This appendix formalizes the limiting case where ESPF recovers UKF behavior, establishing ESPF as a strict generalization.
    
    \subsection{Conditions for Reduction}
    
    The reduction holds under the following conditions:
    \begin{itemize}
        \item All possibility distributions are Gaussian-shaped and interpreted as probabilistic densities.
        \item The initial belief is \( p(x_{k-1}) = \mathcal{N}(\mu_{k-1}, \Sigma_{k-1}) \).
        \item The measurement noise is Gaussian with known covariance \( R \), and additive: \( y_k = h(x_k) + v_k \).
        \item Fusion is performed via multiplication (Bayes' rule), and all distributions are normalized.
        \item sigma-points are generated via the UKF's unscented transform (scaled and symmetrically placed around the mean).
    \end{itemize}
    
    \subsection{Prediction Step Equivalence}

    In the ESPF framework, the prediction step propagates a set of support-points \( \{ \chi^{(i)}_{k-1} \} \) through the nonlinear process model:
    \[
    \chi^{(i)}_{k|k-1} = f(\chi^{(i)}_{k-1})
    \]
    These support-points are generated from a possibilistic kernel centered at the mode \( \hat{x}_{k-1} \), with geometric spread characterized by the plausibility spread matrix \( \Pi_{k-1} \). In the Gaussian limit, this kernel has the form:
    \[
    \pi_{x_{k-1}}(x) \propto \exp\left( -\frac{1}{2}(x - \mu_{k-1})^\top \Sigma_{k-1}^{-1} (x - \mu_{k-1}) \right)
    \]
    where \( \mu_{k-1} = \hat{x}_{k-1} \) and \( \Sigma_{k-1} = \Pi_{k-1} \) under the probabilistic reinterpretation.
    
    The propagated support-points \( \chi^{(i)}_{k|k-1} \) are then used to compute the predicted mean and covariance:
    \begin{align*}
    \mu_{k|k-1} &= \sum_{i=0}^{2n} w_m^{(i)} \chi^{(i)}_{k|k-1} \\
    \Sigma_{k|k-1} &= \sum_{i=0}^{2n} w_c^{(i)} \left( \chi^{(i)}_{k|k-1} - \mu_{k|k-1} \right) \left( \chi^{(i)}_{k|k-1} - \mu_{k|k-1} \right)^\top + Q
    \end{align*}
    where \( w_m^{(i)} \), \( w_c^{(i)} \) are the support-point weights for mean and covariance respectively, and \( Q \) is the process noise covariance.
    
    This procedure yields a predicted Gaussian-shaped kernel centered at \( \mu_{k|k-1} \), with spread matrix \( \Sigma_{k|k-1} \). In the probabilistic interpretation, this becomes the predicted prior:
    \[
    p(x_{k|k-1}) = \mathcal{N}(\mu_{k|k-1}, \Sigma_{k|k-1})
    \]
    
    \subsubsection{Equivalence Statement.}
    If the ESPF kernel is interpreted as a probability density function and all support-points follow the unscented transform rule, the ESPF prediction step is mathematically identical to the UKF prediction:
    \[
    \text{ESPF}_{\text{predict}} \xrightarrow[\text{Gaussian + additive noise}]{\text{probabilistic semantics}} \text{UKF}_{\text{predict}}
    \]
    
    This shows that the ESPF prediction reduces to the UKF prediction under Gaussian assumptions, validating that ESPF recovers classical behavior in the appropriate limit while remaining general enough to handle non-Gaussian, bounded, or even adversarial epistemic conditions.

    \subsection{Measurement Update Equivalence}

    The ESPF measurement update is defined possibilistically through a pointwise \texttt{min} operation:
    \[
    \pi_{x_k|\mathbf{y}_k}(x) = \min \left( \pi_{x_k}(x), \pi_v(y_k - h(x)) \right)
    \]
    where \( \pi_{x_k}(x) \) is the predicted state possibility distribution and \( \pi_v(\cdot) \) encodes the epistemic spread of measurement noise.
    
    In the limiting case where:
    \begin{itemize}
        \item All distributions are Gaussian-shaped and interpreted probabilistically;
        \item The predicted state is \( p(x_k) = \mathcal{N}(\mu_{k|k-1}, \Sigma_{k|k-1}) \);
        \item The measurement noise is additive and Gaussian: \( v_k \sim \mathcal{N}(0, R) \);
        \item Fusion is done multiplicatively (Bayes' rule);
    \end{itemize}
    the ESPF measurement update becomes the standard Bayesian update:
    \[
    p(x_k | y_k) \propto p(x_k) \cdot p(y_k | x_k)
    \]
    with the likelihood given by:
    \[
    p(y_k | x_k) = \mathcal{N}(h(x_k), R)
    \]
    
    To implement this update, the UKF propagates the support-points \( \{ \chi^{(i)}_{k|k-1} \} \) through the nonlinear measurement function:
    \[
    \gamma_k^{(i)} = h(\chi^{(i)}_{k|k-1})
    \]
    The predicted measurement mean and covariance are then estimated as:
    \begin{align*}
    \hat{y}_k &= \sum_{i=0}^{2n} w_m^{(i)} \gamma_k^{(i)} \\
    S_k &= \sum_{i=0}^{2n} w_c^{(i)} \left( \gamma_k^{(i)} - \hat{y}_k \right)\left( \gamma_k^{(i)} - \hat{y}_k \right)^\top + R
    \end{align*}
    The cross-covariance between the state and the measurement is:
    \[
    \Sigma_{xy} = \sum_{i=0}^{2n} w_c^{(i)} \left( \chi^{(i)}_{k|k-1} - \mu_{k|k-1} \right)\left( \gamma_k^{(i)} - \hat{y}_k \right)^\top
    \]
    Using these, the Kalman gain is computed:
    \[
    K_k = \Sigma_{xy} S_k^{-1}
    \]
    The posterior mean and covariance are then updated using:
    \begin{align*}
    \mu_{k|k} &= \mu_{k|k-1} + K_k (y_k - \hat{y}_k) \\
    \Sigma_{k|k} &= \Sigma_{k|k-1} - K_k S_k K_k^\top
    \end{align*}
    
    \subsubsection{Equivalence Statement.}
    If all possibility distributions are interpreted as probabilistic densities, and if support-points and weights follow the Unscented Transform rules, then the ESPF measurement update reduces exactly to the UKF update step:
    \[
    \text{ESPF}_{\text{update}} \xrightarrow[\text{Gaussian + additive noise}]{\text{probabilistic semantics}} \text{UKF}_{\text{update}}
    \]

    \subsection{Interpretation}

    Thus, when ESPF operates under the following assumptions:
    \begin{itemize}
        \item The prior and noise models are Gaussian-shaped and interpreted as normalized probability densities;
        \item The process and measurement models are differentiable and nonlinear;
        \item The fusion step is performed via multiplication (as in Bayes' rule) instead of the possibilistic \texttt{min} operator;
        \item support-points are generated using the standard unscented transform with appropriate weights;
    \end{itemize}
    then the behavior of ESPF becomes mathematically equivalent to the Unscented Kalman Filter (UKF).
    
    In this limiting case:
    \begin{itemize}
        \item The ESPF's possibilistic mode becomes the UKF's probabilistic mean;
        \item The plausibility spread matrix \( \Pi_k \) becomes the state covariance matrix \( \Sigma_k \);
        \item The sparse grid or epistemic sampling collapses to the standard UKF support-point set;
        \item The measurement update via sup--min fusion becomes the Kalman correction step.
    \end{itemize}
    
    \textbf{Conclusion:} The Epistemic Support-Point Filter (ESPF) is a \emph{strict generalization} of the Unscented Kalman Filter. It inherits the structure and recursive flow of UKF but relaxes its core assumptions, replacing:
    \begin{itemize}
        \item Probability with possibility,
        \item Mean-centric statistics with mode-centric epistemic support,
        \item Gaussian constraints with flexible bounded support structures,
        \item Integration with sparse grid maxitivity,
        \item Assumed noise distributions with robust plausibility envelopes.
    \end{itemize}
    
    This allows ESPF to operate reliably in regimes where:
    \begin{itemize}
        \item Knowledge is incomplete or vague;
        \item Evidence is bounded, non-Gaussian, or adversarially perturbed;
        \item Epistemic falsifiability is preferred over statistical optimality.
    \end{itemize}
    
    Yet, when the probabilistic assumptions of the UKF are met, ESPF reduces seamlessly to the UKF, ensuring backward compatibility with classical filtering while expanding its domain of epistemic validity.

\end{document}